\documentclass{raa}
\usepackage{graphicx,times}
\usepackage{natbib}
\usepackage{amssymb,amsmath}
\usepackage{gensymb}
\bibpunct{(}{)}{;}{a}{}{,}

\defcitealias{Bizyaev2014}{B14}
\defcitealias{Bizyaev2009}{BM09}
\defcitealias{Salo2015}{S15}
\defcitealias{Sheth2010}{S10}
\defcitealias{Comeron2018}{C18}
\defcitealias{Mosenkov2010}{M10}
\defcitealias{Mosenkov2018}{M18}
\defcitealias{Mosenkov2021}{M21}
\defcitealias{Bianchi2007}{B07}
\defcitealias{deGeyter2014}{DG14}
\defcitealias{Peters2017}{P17}
\defcitealias{Xilouris1999}{X99}
\defcitealias{DiazGarcia2019}{DG19}
\defcitealias{Pinna2023}{P23}

\usepackage[breaklinks,colorlinks,urlcolor=blue,citecolor=blue,linkcolor=blue]{hyperref}

\begin{document}

\title{Comparing the structural parameters of the Milky Way to other spiral galaxies}

\volnopage{Vol.0 (20xx) No.0, 000--000}
   \setcounter{page}{1}

\author{Jacob A. Guerrette
    \inst{1}
\and Aleksandr V. Mosenkov
    \inst{1}
\and Dallin Spencer
    \inst{1}
\and Zacory D. Shakespear
    \inst{1}
}

\institute{\textsuperscript{1} Department of Physics and Astronomy, N283 ESC, Brigham Young University, Provo, Utah 84602, USA; {\it vinnieguerrette@gmail.com}}

\date{Accepted XXX. Received YYY; in original form ZZZ}



\abstract{The structural parameters of a galaxy can be used to gain insight into its formation and evolution history. In this paper, we strive to compare the Milky Way's structural parameters to other, primarily edge-on, spiral galaxies in order to determine how our Galaxy measures up to the Local Universe. For our comparison, we use the galaxy structural parameters gathered from a variety of literature sources in the optical and near-infrared wavebands. We compare the scale length, scale height, and disk flatness for both the thin and thick disks, the thick-to-thin disk mass ratio, the bulge-to-total luminosity ratio, and the mean pitch angle of the Milky Way's spiral arms to those in other galaxies. We conclude that many of the Milky Way's structural parameters are largely ordinary and typical of spiral galaxies in the Local Universe, though the Galaxy's thick disk appears to be appreciably thinner and less extended than expected from zoom-in cosmological simulations of Milky Way-mass galaxies with a significant contribution of galaxy mergers involving satellite galaxies. 
}

\authorrunning{J. A. Guerrette et al.}
\titlerunning{Milky Way Comparisons}

\maketitle



\section{Introduction}
\label{sec:introduction}

Galaxies are complex conglomerations of stars, gas, dust, and dark matter that can have a variety of different shapes, sizes, and other general properties. Due to this complexity, galaxies can contain a number of different structural components, including active galactic nuclei, bulges, bars, rings, lenses, disks, and stellar halos that determine their morphology \citep{Buta2011}. By studying the structural parameters of galaxies, we get valuable information on their formation history through galaxy scaling relations, as different groupings of galaxies on these scaling relations may point to different scenarios of their formation and evolution (see \citealt{Hohl1978,Combes1981,Michel2004,Knapen2004,Knapen2013}). Furthermore, by studying galaxies that are relatively similar in morphology, we can begin to form an idea of the sequence of events required to create that variety of galaxies \citep{MunozMateos2015}.

Due to our location within one of its spiral arms, the Milky Way, while being our own Galaxy, remains somewhat of a mystery to us. The dust in the interstellar medium (ISM) blocks our view of much of the Galactic plane in addition to scattering what light we do receive (see \citealt{Draine2003} and references therein). Because of this, the specific structural parameters describing our Galaxy are difficult to study.

Luckily, due to infrared-submillimeter-radio observations where our Galaxy is largely transparent, we can explore its stellar, dusty, and gaseous content in great detail. In a review about our Galaxy's parameters, \citet{Bland2016} state, however, that there is much uncertainty about the Galaxy's structural parameters, especially parameters such as disk scale length ($\sim$2$-$4 kpc; see e.g. \citealt{Kent1991,Ruphy1996,Drimmel2001,Lopez2002,Benjamin2005,Reyle2009}) and scale height ($\sim$150$-$375 pc; see e.g. \citealt{Kroupa1992,Gould1996,Ojha2001,Siegel2002,Juric2008}). This uncertainty in the Galaxy's structural parameters makes it difficult to compare our Galaxy with other similar spiral galaxies.

Recently, however, using unWISE 3.4~{\textmu}m integrated photometry \citep{Lang2014}, \citeauthor{Mosenkov2021} (\citeyear{Mosenkov2021}, henceforth \citetalias{Mosenkov2021}) fitted multiple disk models of the Milky Way to determine which model best matches the observational data. Importantly, they used the full treatment of the sky projection and dust extinction effects in their 3D modelling of the Galaxy structure. Another advantage of their work was that their approach of photometric decomposition was essentially the same as has been used for external galaxies, making the comparison of our Galaxy with other galaxies more reliable. They revealed that the best fitting model contained two disks with exponential vertical density profiles, outer flaring, warping, and two over-densities on either side of the Galactic center represented by two blobs (see their table~1). Although this was the best-fitting two-disk model, a similar model (with the only difference being a sech$^2$ vertical density profile instead of an exponential one, henceforth called an isothermal disk) was a close second-best fit. Here, the simple exponential model \citep{1989ApJ...337..163W,Xilouris1999,2011MNRAS.414.2446M,2012MNRAS.426.1328B} has a luminosity density profile $j(R,z)$ given by
\begin{equation}
    j(R,z)=j_0~\text{e}^{-R/h_\mathrm{R}-|z|/h_\mathrm{z}}
    \label{Eq:Exponential}
\end{equation}
where $j_0$ is the central luminosity density of the disk, $h_\mathrm{R}$ is the disk scale length, and $h_\mathrm{z}$ is the disk scale height. Similarly, the isothermal model \citep{1942ApJ....95..329S,1950MNRAS.110..305C,1981A&A....95..105V,1981A&A....95..116V,1982A&A...110...61V} has a luminosity density distribution given by
\begin{equation}
    j(R,z)=j_0~\text{e}^{-R/h_\mathrm{R}}~\text{sech}^2[|z|/(2h_\mathrm{z})].
    \label{Eq:Isothermal}
\end{equation}

Both functional dependencies are often used in the literature to describe the disk structure, but for our Galaxy none of the models provides a significantly better fit. Therefore, throughout this paper, we will simultaneously use both models for comparison purposes.

\begin{figure*}
    \centering
    \includegraphics[width=.55\textwidth,height=6.7cm]{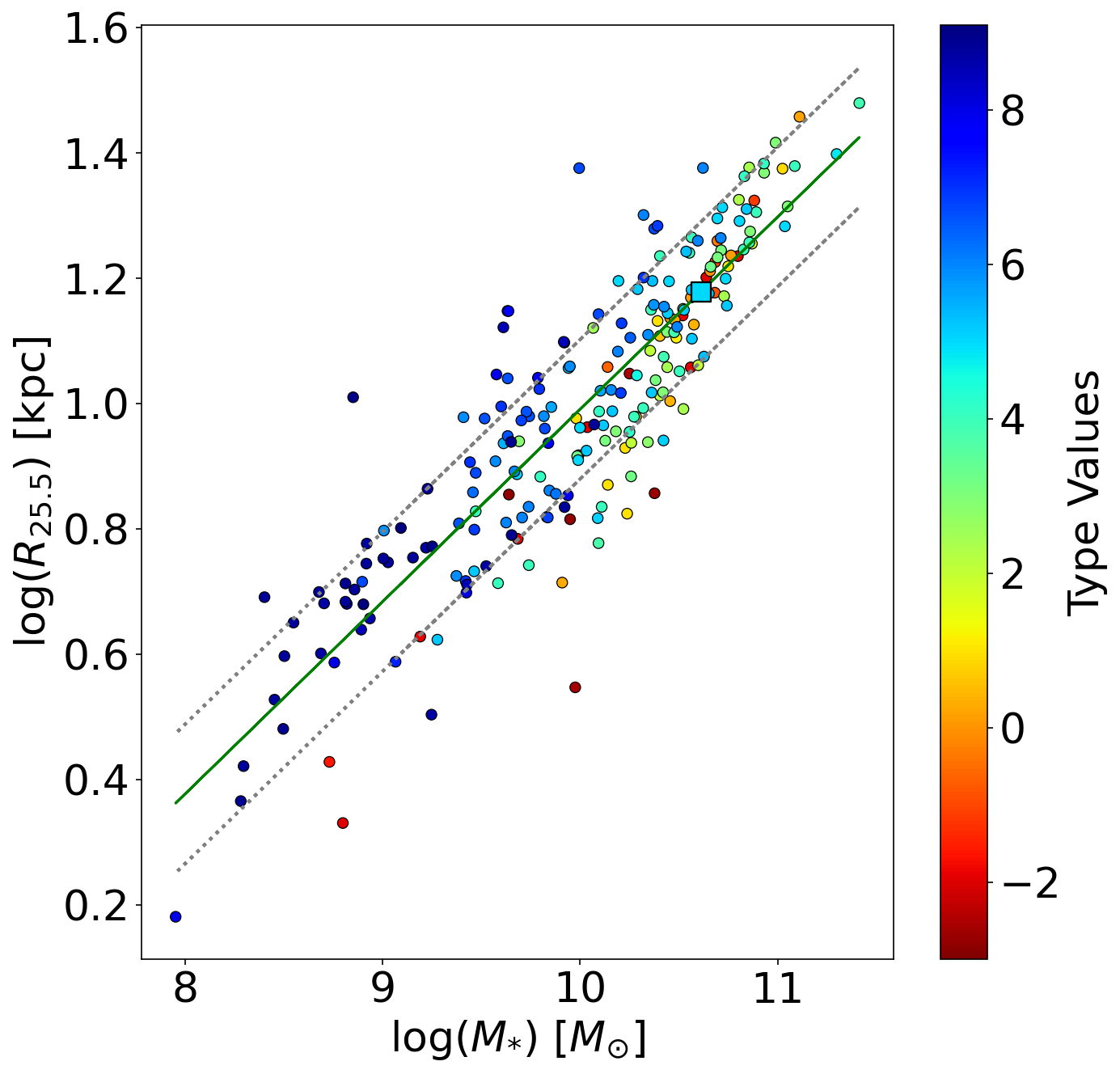}\hfill
    \\[\smallskipamount]
    \caption{Semi-major axis at $\mu_{3.6} = 25.5$ AB mag\,arcsec$^{-2}$ for face-on ($i<40\degr$) disk galaxies from \citetalias{Sheth2010}. Data points are colored based on the galaxy's numerical Hubble stage value. A trendline is included, along with a $\pm 1\sigma$ spread. The Milky Way semi-major axis value of 15.1 kpc was calculated using prescriptions from section 2.1 in \citet{Pilyugin2023} adopting the surface stellar mass density $\Sigma_{*,R_0}=38$~M$_{\odot}$\,pc$^{-2}$ at the solar galactocentric distance $R_0=8.0$~kpc. The Milky Way mass value is taken from \citet{Bland2016}. The Galaxy is denoted by the blue square, corresponding to a Type value of 5.}
    \label{fig:s4g_sma_mass}
\end{figure*}

In this paper, our ambition is to compare some of the most recent Milky Way structural models to those obtained for other spiral galaxies in the Local Universe, helping us conclude how the Milky Way fits into the general picture. This is done through studying the general statistics of spiral galaxies, for which we have structural parameters from the literature, and by exploring general galaxy scaling relations for a large range of stellar masses (see e.g. \citealt{Mosenkov2010,2014MNRAS.441.1066M}) with the Milky Way superimposed. For example, in Figure~\ref{fig:s4g_sma_mass}, one can see that the Milky Way perfectly follows the general trend of the mass--size relation. We utilize one-dimensional, two-dimensional, and three-dimensional modeling to compare the structural parameters of our Galaxy to external edge-on and non-edge-on spiral galaxies. Our goal is to consider models obtained with \textit{the same or similar decomposition methods} for maximum accuracy. By comparing the Milky Way structural models with those for other galaxies, we can determine how normal or abnormal the Milky Way is, which, in turn, can give us a better understanding of its evolutionary status. 

The paper is organized as follows. In Section~\ref{sec:samples}, we present galaxy samples with available photometric decomposition taken from the literature for our comparative analysis. In Section~\ref{sec:optical}, we consider various models of our Galaxy and external galaxies in the optical. In Section~\ref{sec:singleIR}, we detail the single-disk data comparisons in the near-infrared (NIR). In Section~\ref{sec:doubleIR}, we continue detailing data comparisons in the NIR but for the two-disk models. In Section~\ref{sec:bulge}, we discuss how the bulge/bar in our Galaxy compares to the bulges in external galaxies. In Section~\ref{sec:pitch_angle}, we discuss how the mean pitch angle of the Milky Way compares to external galaxies with similar bulge-to-total flux ratios and similar morphological types. Finally, in Section~\ref{sec:conclusion}, we summarize our findings.

\section{The samples}
\label{sec:samples}

The structural parameters of a galaxy can vary with wavelength due to dust attenuation and stellar population gradients (see \citealt{deJong1996, MacArthur2003, Taylor2005, Fathi2010}). Thus, it is important to analyze galaxy models in both optical and NIR wavelengths in order to best determine how typical the Milky Way is among similar galaxies at wavelengths where a mixed stellar population radiates (the optical part of the spectrum) and where old stellar population makes up the bulk of the stellar mass (NIR). For comparison purposes, we only use models obtained by means of the same or similar methods and within similar wavebands.

We employ single-disk and two-disk galaxy models \citep{Salo2015,Comeron2018} and galaxy pitch angles \citep{DiazGarcia2019} from the {\it Spitzer} Survey of Stellar Structure in Galaxies (S$^4$G, \citealt{Sheth2010,MunozMateos2013,2015ApJS..219....5Q}) and single-disk models from \citet{Bizyaev2009} and \citet{Mosenkov2010} based on the Two Micron All Sky Survey (2MASS, \citealt{Skrutskie2006}) as our tools for comparing these NIR models of the Milky Way to other galaxies. For these various sources, we take orientation into account, separately comparing the Milky Way to edge-on (inclinations greater than or equal to roughly 80 degrees; \citealt{Salo2015}) and non-edge-on (inclinations less than 80 degrees) galaxies.

To obtain an accurate comparison in the optical, we also utilize the {\it optical} Milky Way model from \citet{Natale2022}. They created a radiative transfer (RT) model of the Milky Way in the \textit{V} band, from which we extracted the structural parameters of a general stellar disk. For comparison, we gathered a sample of RT models of galaxies in optical wavelengths from four papers: \citet{Bianchi2007}, \citet{deGeyter2014}, \citet{Peters2017}, and \citet{Xilouris1999}. In addition, for our comparative analysis, we exploit $\sim$2800 3D models of edge-on galaxies from \citet{Bizyaev2014}, who used $r$-band data from the Sloan Digital Sky Survey (SDSS, \citealt{York2000,Eisenstein2011}). 

In Table~\ref{tab:ref_table}, we provide brief information on the samples used. Once again, we emphasize that in almost every case, we only compare the structure of our Galaxy in a given waveband to the structure of other galaxies observed within that same or a close waveband. This same restriction applies to the methods used to model these galaxies, as well as to the complexity of the model. By restricting these factors, we hope to isolate the structural parameters of the galaxies we are analyzing. In the next sections, we make a comparison of the Milky Way with other galaxies in the optical and NIR regions of the spectrum using the aforementioned data sets. 

Below we provide brief summaries of the data samples used, including details on the modelling techniques employed.

\citet{Bianchi2007} used RT modeling to retrieve the structural parameters of 7 galaxies in the \textit{V} and \textit{K'} bands. They performed their own observations at the 3.5-m TNG (Telescopio Nazionale Galileo) telescope located at the Roque de Los Muchachos Observatory in La Palma, Canary Islands. Their method involved the iterative production of a model image of a galaxy which is then compared with the observed image in an attempt to derive the structural parameters that minimize $\chi^2$ (see their section 3 for an in-depth analysis of their methods).

\citet{Bizyaev2009} analyzed 139 galaxies from the Revised Flat Galaxies Catalog (RFGC, \citealt{Karachentsev1999}) using 2MASS images \citep{Skrutskie2006} in the \textit{J}, \textit{H}, and \textit{K}$_s$ wavebands. To obtain the structural parameters of these galaxies, they used the same method described in section 3 of \citet{Bizyaev2002} which is based on analyzing photometric profiles drawn parallel to the major and minor axes of a galaxy at a one-pixel interval (see their section 2 for further details).

\citet{Bizyaev2014} investigated the structural parameters of 5747 genuine edge-on galaxies that were suitable for automatic analysis from SDSS \citep{York2000}, RFGC \citep{Karachentsev1999}, the Third Reference Catalog of Bright Galaxies (RC3, \citealt{deVaucouleurs1991}), the EFIGI catalogue \citep{Baillard2011}, and GalaxyZoo \citep{Lintott2011} in the \textit{g}, \textit{r}, and \textit{i} wavebands. Of these 5747 galaxies, 3912 were fitted using a 3D model by estimating a set of 1D galaxy photometric profiles and accounting for dust absorption. We utilized 2789 galaxies from this sample that had radial velocity data 
 to perform redshift calculations and convert the structural parameters to kiloparsecs.

\citet{Comeron2018} examined 141 galaxies using images from S$^4$G and its early-type extension \citep{Sheth2010,Sheth2013} in the 3.6 micron waveband. Their complex fitting procedure involved steps the following steps: 1) fitting the vertical surface-brightness profiles of galaxies (while accounting for the gravitational effect of the gas disk), 2) fitting the axial surface-brightness profiles, 3) refitting the vertical profiles while accounting for the presence of central mass concentration (CMC) light, 4) producing an axial surface-brightness profile for heights dominated by the thin and thick disks, and 5) calculating the mass of each component (see their section 3 for an in-depth breakdown of their methods).

\citet{deGeyter2014} analyzed 12 galaxies from the Calar Alto Legacy Integral Field Area survey (CALIFA, \citealt{Sanchez2012}) in the \textit{g}, \textit{r}, \textit{i}, and \textit{z} wavebands. To create RT models of these galaxies, they used the FitSKIRT automated fitting routine \citep{deGeyter2013} which couples the SKIRT Monte Carlo RT code \citep{Baes2011} with GAlib, the genetic algorithms-based optimization library. FitSKIRT takes an arbitrary number of images and fits a RT model with an arbitrary number of dust/stellar components (see their section 3 for further information on their modeling).

\citet{DiazGarcia2019} studied 391 spiral galaxies from S$^4$G \citep{Sheth2010} in the 3.6 micron waveband. Of these 391 galaxies, we use models for 233 of them that are classified as multi-armed or grand design. To model these galaxies, \citet{DiazGarcia2019} used deprojected images of the galaxies to calculate global pitch angles from average measurements of individual logarithmic spiral segments. Additionally, for a small subsample of galaxies with measurements available in the literature, they performed a 2D Fourier analysis by using Fourier transform spectral analyses of the deprojected galaxy images (see their section 3 for more details on these methods).

\citet{Mosenkov2010} investigated the structural parameters of 175 edge-on galaxies in the \textit{K}$_s$ waveband selected from the 2MASS-selected Flat Galaxy Catalog (2MFGC, \citealt{Mitronova2004}). They used the BUDDA decomposition code \citep{deSouza2004} to fit the parameters of the S\'ersic bulge and the edge-on disk. BUDDA takes into account the entire 2D image of a galaxy to find the optimal galaxy model by employing the multidimensional downhill simplex method \citep{Press1989}.

\citet{Mosenkov2018} examined 7 large edge-on galaxies from the \textit{Herschel} \citep{Pilbratt2010} Observations of Edge-on Spirals (\textit{HER}OES) sample. They used available databases for optical and NIR sky surveys to gather images of these galaxies, then used a three step process to model the galaxies: first, they fit a general model to each galaxy. Then, they used FitSKIRT to fit a dust model to each galaxy. Finally, they performed SKIRT panchromatic radiative transfer simulations from their FitSKIRT models to obtain the final models.

\citet{Peters2017} used their own observations from 10 telescopes in their modeling and analysis of 8 galaxies (with one galaxy being split into left and right halves) in the \textit{B}, \textit{V}, \textit{R}, \textit{I}, \textit{J}, \textit{H}, \textit{K}, Wise1, and Wise2 wavebands. Similar to \citet{deGeyter2014}, they utilized the FitSKIRT automated fitting routine to model the structural parameters of their galaxies.

\citet{Pinna2023} studied 24 galaxies from the AURIGA project\footnote{https://wwwmpa.mpa-garching.mpg.de/auriga/} \citep{Grand2017}, which performs zoom-in magneto-hydrodynamical cosmological simulations of 30 Milky Way-mass galaxies. They projected these galaxies in an edge-on view, then spatially binned them using Voronoi binning \citep{Cappellari2003}. Finally, they decomposed their edge-on galaxies into two disk components by fitting vertical-brightness profiles to the galaxies to obtain their models.

\citet{Sheth2010} analyze 2331 galaxies observed with the \textit{Spitzer} space telescope \citep{Werner2004} at 3.6 and 4.5{\textmu}m. While their method for obtaining models is complex and best described by their section 5, the summary is as follows: for each galaxy, they use the initial observations to gather parameters such as diameter and axial ratio. Then, they measure things like total magnitude and the azimuthally-averaged radial profile of the surface brightness. Finally, they perform a decomposition using \textsc{galfit} to obtain the main constituent stellar components of each galaxy.

\citet{Salo2015} examined 2352 galaxies from S$^4$G \citep{Sheth2010} in the 3.6 and 4.5 micron wavebands. To model these galaxies, they performed 2D surface brightness structural decompositions using \textsc{galfit3.0} \citep{Peng2010} which uses the Levenberg-Marquadr algorithm to minimize the weighted residual $\chi_{\nu}^2$ between the observations and the models. Additionally, they performed human-supervised multi-component decompositions for each galaxy (see their section 2 for their decomposition pipeline).

Finally, \citet{Xilouris1999} used their own observations to create RT models for 5 galaxies in the \textit{B}, \textit{V}, \textit{I}, \textit{J}, and \textit{K} wavebands. These RT models were based on surface photometry (using the same process as \citealt{Kylafis1987}) and were used to input model parameters into their equations 1-6 to find the parameters that best match the observed surface photometry of each galaxy.

\begin{table}
	\centering
	\caption{Details on the various data samples utilized. For subsamples from the same source paper, a number in square brackets is added after their label to differentiate between different subsamples. Details on why we only use a specific number of galaxies from a paper are given in the caption.}
	\label{tab:ref_table}
    \resizebox{\textwidth}{!}
    {\begin{tabular}{l l l l l} 
	    \hline\\[-0.5ex]
		Label & Method & Reference & \# of Galaxies & Waveband\\
		&of fitting&&&used
		\\[0.5ex]
	    	\hline\\[-0.5ex]
		
		B07 & RT & \citet{Bianchi2007} & 7 & \textit{V}\\

	    BM09 & 1D & \citet{Bizyaev2009} & 139 & \textit{K$_\mathrm{s}$}\\
		
	    B14 & 3D & \citet{Bizyaev2014} & 2789 & \textit{r}\\

	    C18\textsuperscript{1} [1] [2] & 1D & \citet{Comeron2018} & 
        96, 135 & 3.6 {\textmu}m \\

	    DG14 & RT & \citet{deGeyter2014} & 12 & $g, r, i, z$\\

        DG19 & 2D & \citet{DiazGarcia2019} & 233 & 3.6 {\textmu}m\\

	    M10 & 2D & \citet{Mosenkov2010} & 175 & \textit{K$_\mathrm{s}$}\\

        M18 & RT & \citet{Mosenkov2018} & 7 & Several Optical-NIR bands\\
     
	    P17 & RT & \citet{Peters2017} & 9 & Several UV-MIR bands\\

        P23 & 1D & \citet{Pinna2023} & 24 & \textit{V}\\

        S10 & 2D & \citet{Sheth2010} & 222 & 3.6 {\textmu}m\\
     
	    S15\textsuperscript{2} [1] [2] [3] & 2D & \citet{Salo2015} & 1287, 217, 128, & 3.6 {\textmu}m\\
	    
	    [4] [5] &  &  & 115, 576 & \\

        X99 & RT & \citet{Xilouris1999} & 5 & \textit{V} \\[+0.5ex]
		\hline
    \end{tabular}}
    \begin{minipage}{15cm}
    \vspace{0.1cm}
    \footnotesize
    {\textsuperscript{1}C18 [1] comprises edge-on two-disk galaxies; [2] consists of galaxies with available disk mass data.\\
    \textsuperscript{2}S15 [1] comprises non-edge-on single-disk galaxies; [2] consists of edge-on single-disk galaxies; [3] includes edge-on two-disk galaxies; [4] contains galaxies 
    with disk mass data; [5] is galaxies with B/T ratio data.}
    \end{minipage}
\end{table}

\section{The single-disk models in the optical}
\label{sec:optical}

\begin{table}
	\centering
	\caption{MW model structural parameter values from various sources. Scale length, scale height, and radius values are given in kpc, the stellar masses are given in solar masses, ratio values are unitless, magnitudes are given in mag, and the absolute value of the mean pitch angle is given in degrees.}
	\label{tab:MW_table}
    \resizebox{\textwidth}{!}
    {\begin{tabular}{l l l l}
	    \hline\\[-0.5ex]
		Model Type & Reference & Parameters & Values\\[+0.5ex]
	    \hline\\[-0.5ex]
		
		RT (Optical) 1-disk & \citet{Natale2022} & $h_\mathrm{R}$, $h_\mathrm{z}$, Ratio & 3.10, 0.30, 10.33
      \\[+0.8ex]

	    NIR 1-disk Exp. & \citetalias{Mosenkov2021} & $h_\mathrm{R}$, $h_\mathrm{z}$, Ratio & 3.19, 0.45, 7.02
     \\[+0.8ex]

        NIR 1-disk Iso. & \citetalias{Mosenkov2021} & $h_\mathrm{R}$, $h_\mathrm{z}$, Ratio & 3.02, 0.48, 6.36
     \\[+0.8ex]

	    NIR 2-disk Thick Exp. & \citetalias{Mosenkov2021} & $h_\mathrm{R}$, $h_\mathrm{z}$, Ratio & 3.22, 0.71, 4.55
     \\[+0.8ex]

	    NIR 2-disk Thick Iso. & \citetalias{Mosenkov2021} & $h_\mathrm{R}$, $h_\mathrm{z}$, Ratio & 2.79, 0.98, 2.85
     \\[+0.8ex]

	    NIR 2-disk Thin Exp. & \citetalias{Mosenkov2021} & $h_\mathrm{R}$, $h_\mathrm{z}$, Ratio & 2.55, 0.25, 10.11
     \\[+0.8ex]

        NIR 2-disk Thin Iso. & \citetalias{Mosenkov2021} & $h_\mathrm{R}$, $h_\mathrm{z}$, Ratio & 2.52, 0.24, 10.57
     \\[+0.8ex]

        Optical & \citet{Pilyugin2023} & $M_*$, $R_{\mathrm{opt}}$ & $5.2 \times 10^{10}$, 12.0
     \\[+0.8ex]

        Optical & \citet{Bland2016} & \textit{r}-band Abs. Mag. & -21.64
     \\[+0.8ex]
     
	    NIR 2-disk & \citet{Bland2016} & $M_*$ & $4.1 \times 10^{10}$
     \\[+0.8ex]

        NIR 2-disk\textsuperscript{1} & \citetalias{Mosenkov2021} & $B/T$ Max, $M_{\mathrm{T}}/M_{\mathrm{t}}$, $M_{\mathrm{T}}/M_{\mathrm{t}}$ & 0.12, 0.42, 1.34
     \\[+0.8ex]

        Spiral Structure & \citet{Vallee2015} & $|\phi|$ & 13.1
     \\[+0.8ex]
     \hline
    \end{tabular}}
    \begin{minipage}{15cm}
    \vspace{0.1cm}
    \footnotesize
    \textsuperscript{1}Two values are included for the disk mass ratio due to uncertainty in the local thick-to-thin disk surface density ratio (see Sect.
    
    ~\ref{subsec:disk_mass} for further explanation).
    \end{minipage}
\end{table}

\begin{table}
	\centering
        \tiny
	\caption{Quartiles for each model group. For model groups with multiple parameters, the order of the quartile data matches the order of the listed parameters. $h_\mathrm{R}$ and $h_\mathrm{z}$ values are in units of kpc, while ratio values are unitless.}
	\label{tab:quartile_table}
    \resizebox{\textwidth}{!}
    {\begin{tabular}{l l l}
	    \hline\\[-0.5ex]
		Model Type & Parameters & Quartiles\\[+0.5ex]
	    \hline\\[-0.5ex]
		
		Optical 1-disk & $h_\mathrm{R}$, $h_\mathrm{z}$, Ratio &   $4.88^{+2.10}_{-1.62}$, $0.44^{+0.19}_{-0.15}$, $10.68^{+4.83}_{-3.09}$  
      \\[+0.8ex]

	    NIR 1-disk Non-Edge-On & $h_\mathrm{R}$ & $2.38^{+0.88}_{-0.75}$
     \\[+0.8ex]

	    NIR 1-disk Edge-On & $h_\mathrm{R}$, $h_\mathrm{z}$, Ratio & $3.75^{+1.43}_{-1.39}$, $0.43^{+0.17}_{-0.15}$, $8.52^{+2.28}_{-2.21}$
     \\[+0.8ex]

	    NIR 2-disk Thick & $h_\mathrm{R}$, $h_\mathrm{z}$, Ratio & $3.07^{+0.84}_{-0.81}$, $0.60^{+0.46}_{-0.15}$, $4.83^{+2.23}_{-1.58}$
     \\[+0.8ex]

	    NIR 2-disk Thin & $h_\mathrm{R}$, $h_\mathrm{z}$, Ratio &  $1.38^{+0.77}_{-0.49}$, $0.17^{+0.07}_{-0.04}$, $9.22^{+4.48}_{-3.28}$
     \\[+0.8ex]
	    
	    NIR 2-disk & Mass Ratio & $0.95^{+1.07}_{-0.43}$
     \\[+0.8ex]

	    NIR Edge-On & $B/T$ & $0.47^{+0.15}_{-0.25}$
     \\[+0.8ex]

	    NIR Non-Edge-On & $B/T$ & $0.14^{+0.16}_{-0.08}$
     \\[+0.8ex]
     \hline
    \end{tabular}}
\end{table}

We begin by comparing models obtained in optical wavelengths primarily using RT modeling techniques. RT models simulate how photons within the galaxy propagate and interact with the galactic interstellar medium. \citet{Natale2022} (from which we gathered our RT Milky Way parameter values) created images from such RT simulations to compare them to the observed surface photometry of the Milky Way in order to find the most accurate RT model. We gathered the RT model parameters for the stellar disks of other galaxies from \citetalias{Xilouris1999}, \citetalias{Bianchi2007}, \citetalias{deGeyter2014}, and \citetalias{Peters2017} (see Table~\ref{tab:ref_table} for labels, information, and references on any papers used in the making of this paper's figures). In total, we have 33 RT models for external galaxies, which cannot by any means be a representative sample of spiral galaxies. Unfortunately, detailed RT modeling is very time consuming (see e.g. details in \citealt{Mosenkov2018}), so, to our best knowledge, this is the largest combined sample of galaxies with available RT models in the optical. 

Here, we also use models from \citetalias{Bizyaev2014}, who exploited optical observations from the Edge-on Galaxies in the SDSS (EGIS) catalogue. They performed both 1D and 3D photometric decomposition, but in this paper we only utilize their $r$-band 3D models that account for dust attenuation. 

Although we typically cannot study the Milky Way in optical wavelengths due to the dust within the Galactic plane, the \citet{Natale2022} RT model allows one to indirectly estimate our Galaxy's radial scale length through a scaling factor, making their model invaluable to our comparisons within optical wavebands. Their scaling factor comes from table~E.1 of \citet{Popescu2011} who performed RT calculations for a number of spiral galaxies to find a consistent model. They found that the disk scale length varies with wavelength, so they normalized the scale lengths in their table~E.1 to the \textit{B}-band scale length. \citet{Natale2022} then use these normalization factors to estimate the \textit{V}-band scale length of the Milky Way based on the \textit{K}-band scale length they constrained from their data, giving a value of 3.10~kpc. \citet{Popescu2011} also found that the disk scale height does not vary with wavelength, so \citet{Natale2022} utilize their own constrained \textit{K}-band scale height of 0.30 kpc for their \textit{V}-band scale height. 

\begin{figure*}
    \centering
    \includegraphics[width=.33\textwidth, height=3.2cm]{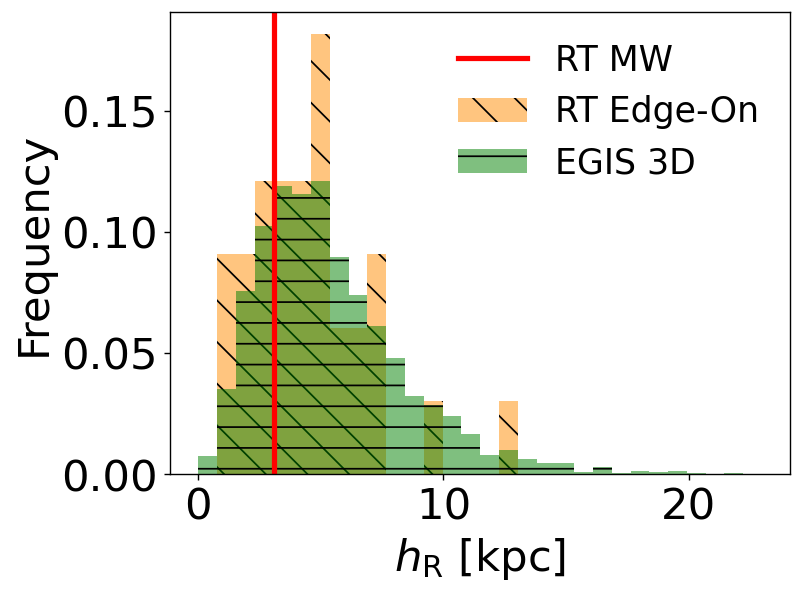}\hfill
    \includegraphics[width=.33\textwidth, height=3.2cm]{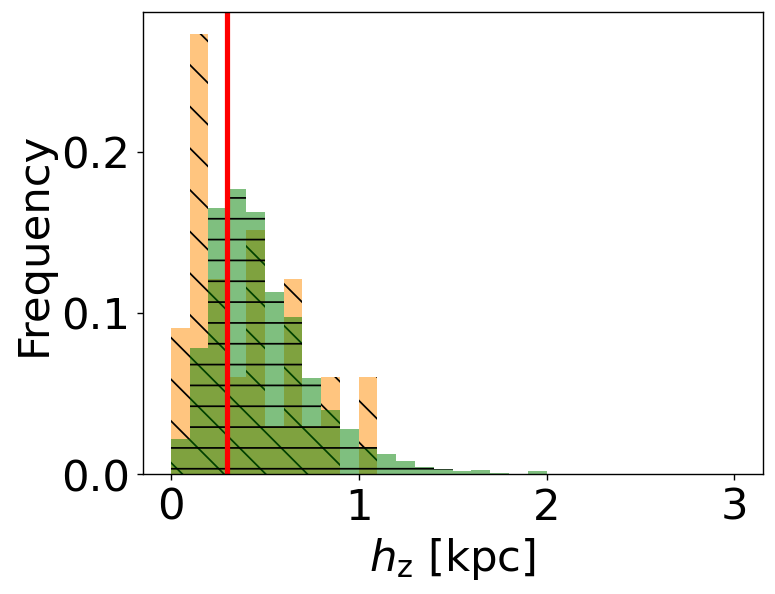}\hfill
    \includegraphics[width=.33\textwidth, height=3.2cm]{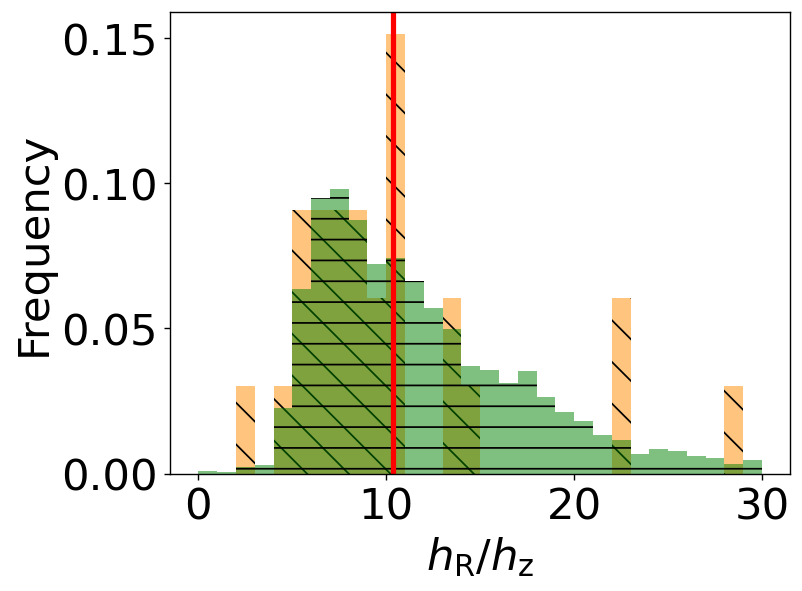}
    \\[\smallskipamount]
    \caption{Histograms of the scale length (\textit{left}), scale height (\textit{middle}), and ratio between the two (\textit{right}), with the respective Milky Way model parameter values from \citet{Natale2022} highlighted. Edge-on single-disk galaxy models in the optical are taken from \citetalias{Xilouris1999, Bianchi2007, deGeyter2014, Peters2017} (`RT Edge-On') and EGIS (3D, \citetalias{Bizyaev2014}).  }\label{fig:multi_optical_graphs}
\end{figure*}

In Figure~\ref{fig:multi_optical_graphs}, we depict the distributions of the selected galaxies by the disk scale length, disk scale height, and the ratio of the two. We point out that the stellar parameters in \citetalias{Xilouris1999} and \citetalias{Bianchi2007} are provided in different wavebands (we use the $V$ band here), while \citetalias{deGeyter2014} and \citetalias{Peters2017} provide the global properties of the stellar fits based on {\it all} wavebands utilized. Despite this difference in the output parameters, we see that the general distributions for the RT and EGIS models are roughly consistent among the datasets under consideration, although the low number of RT data points creates very spiky distributions.

\begin{figure*}
    \centering
    \includegraphics[width=.48\textwidth,height=6.5cm]{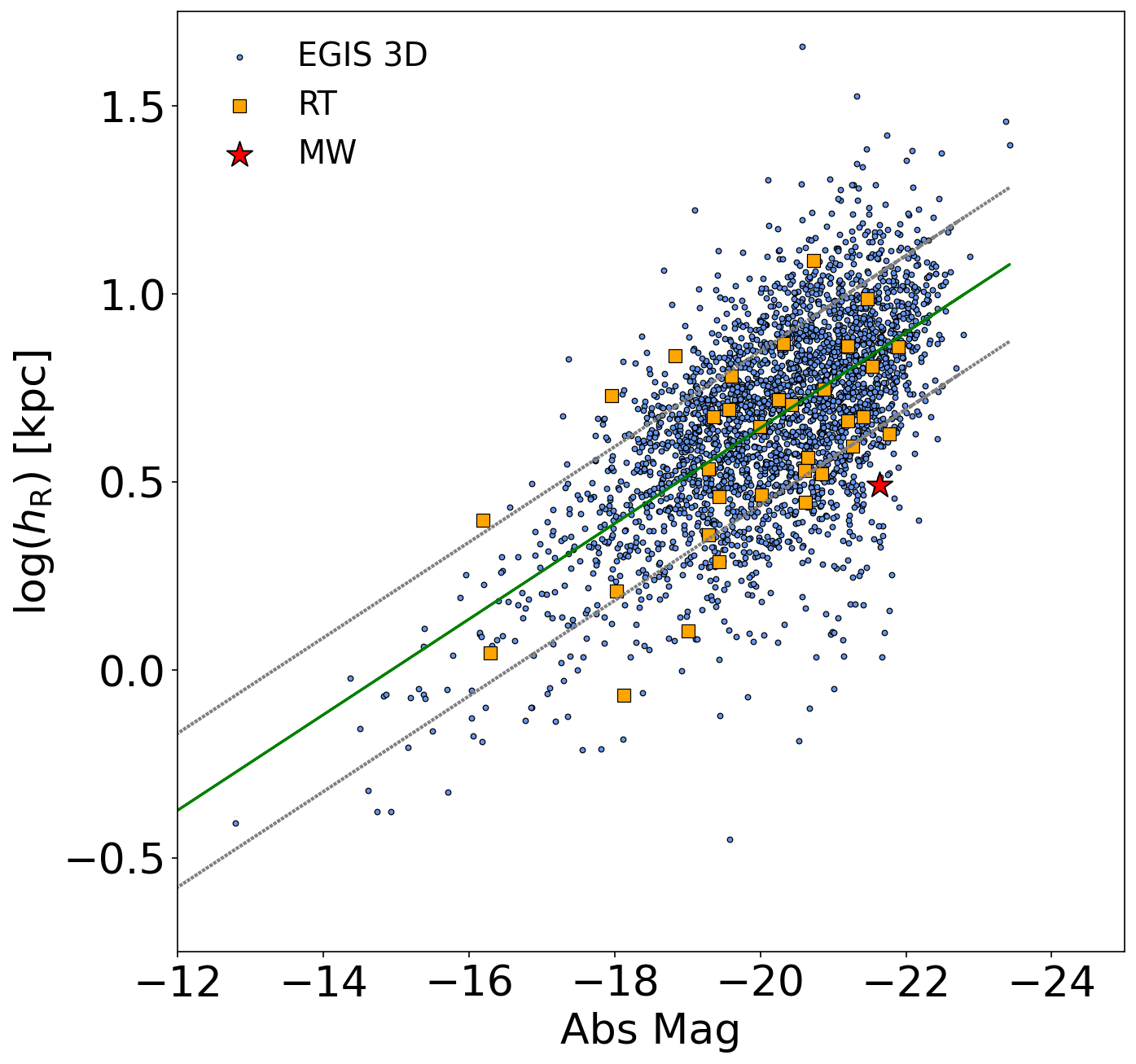}\hfill
    \includegraphics[width=.48\textwidth,height=6.5cm]{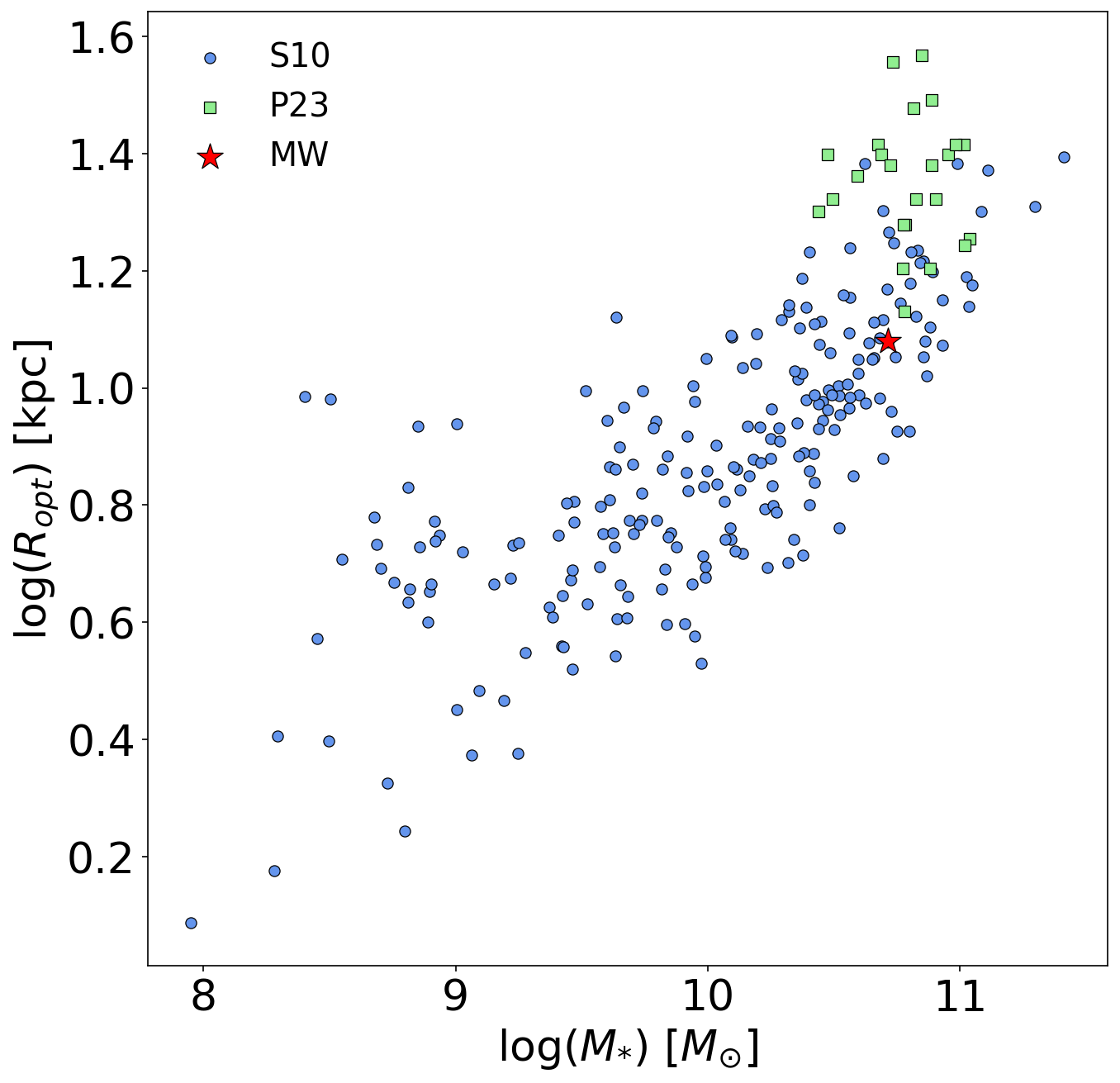}\hfill
    \\[\smallskipamount]
    \caption{\textit{Left}: Absolute r- or R-band magnitude versus scale length for our optical datasets: \citetalias{Bizyaev2014} ('EGIS 3D') and \citetalias{Xilouris1999,Bianchi2007,deGeyter2014,Peters2017} ('RT'). A trendline is included, along with a $\pm 1\sigma$ spread. Milky Way model magnitude is from \citet{Bland2016} and its scale length is from \citet{Natale2022}. The Milky Way data point is denoted by the red star. \textit{Right}: Optical radius (semi-major axis of the 25 mag\,arcsec$^{-2}$ isophote in the $B$ band taken from HyperLeda) versus stellar mass for face-on ($i<40\degr$) disk galaxies from \citetalias{Sheth2010} and 24 Milky Way-mass galaxies from the AURIGA zoom-in cosmological simulations from \citet{Pinna2023}. Milky Way values are from \citet{Pilyugin2023}, with the data point being marked by the red star.}\label{fig:optical_Ropt_mass}
\end{figure*}

Using the combined data in the optical, we can compare the structural parameters of the Milky Way (see Table~\ref{tab:MW_table} for all main Milky Way parameter values) to those for external disk galaxies. From our dataset's quartiles Q1 (25\%) and Q2 (median) scale length cutoff values (see Table~\ref{tab:quartile_table} for this and all future quartile cutoff values), we see that the Milky Way's scale length is lower than average, falling just below the Q1 cutoff. On the other hand, the Q1 and median scale height cutoff values tell us that the Milky Way's scale height is average (close to Q2).

When we compare the flatness of the Milky Way to other galaxies (the \textit{right} plot in Fig.~\ref{fig:multi_optical_graphs}, the scale length divided by the scale height), the Milky Way's value of 10.33 seems to fall around the median value. Indeed, the dataset's Q1 and median cutoffs indicate that the Milky Way's ratio value is close to Q2, making it quite average.

We thus conclude that the Milky Way's scale length is slightly below average in optical bands, although its scale height and flatness ratio are both normal. This conclusion is also confirmed by the deviation of the Milky Way from the general trend on the luminosity--scale length relation --- our Galaxy's disk scale length appears to be appreciably smaller compared to those of disk galaxies with similar luminosities (see Fig.~\ref{fig:optical_Ropt_mass}, {\it left} plot). The optical radius $R_\mathrm{opt}=12$~kpc of our Galaxy, estimated in \citet{Pilyugin2023}, is however, in good agreement with those of disk galaxies with similar stellar masses. However, as one can see from Figure~\ref{fig:optical_Ropt_mass} ({\it right} plot), Milky-Way-like galaxies with similar masses from AURIGA zoom-in cosmological simulations \citep{Grand2017} have significantly larger (0.3 dex) optical radii. We will return to this fact in Section~\ref{sec:conclusion}.

\section{The single-disk models in the NIR}
\label{sec:singleIR}

In this section, we present our analysis of single-disk galaxy models created in NIR wavelengths, specifically the \textit{K$_\mathrm{s}$} and 3.6~$\mu$m bands. As a note, we neglect the effects of dust attenuation and polycyclic aromatic hydrocarbon emission in the case of the 3.6~$\mu$m waveband in these models (see \citetalias{Mosenkov2021}). In Section~\ref{sec:optical} where we explore edge-on disk models in the optical, all galaxy models available in the literature consist of a single disk. Fitting two-disk models in the optical is problematic because of the strong dust attenuation in the plane of the galaxy where a thin disk may be present. In the NIR, the observed vertical structure of an edge-on galaxy is much less distorted by the internal extinction, so a large number of both single-disk and two-disk models of edge-on galaxies have been fitted in the literature. Therefore, we differentiate these one-disk and two-disk models and describe them in separate sections. In this section, we focus on the NIR single-disk models (including those fitted for non-edge-on galaxies for comparison), whereas the NIR two-disk models are analyzed separately in Section~\ref{sec:doubleIR}. 

As detailed in Section~\ref{sec:introduction}, for our Milky Way values, we make use of two single-disk models from \citetalias{Mosenkov2021} obtained in 3.4~$\mu$m: one with a single {\it exponential} disk and another with a single {\it isothermal} disk, both with outer flaring. Having two points of reference for the Milky Way is beneficial in some instances, as when both are out of the ordinary, it is indicative that the Milky Way might have a non-typical structure. Also, it is known from the literature that some vertical galaxy profiles are better fit with an exponential disk, and some with an isothermal disk \citep[see e.g.][and references therein]{2011ARA&A..49..301V}. For our Galaxy, \citetalias{Mosenkov2021} find that a single isothermal disk slightly better agrees with the 3.4~$\mu$m Galaxy composite image than a similar model but with an exponential disk. However, as noted in Section~\ref{sec:introduction}, the difference between the exponential and isothermal disk models is not significant enough to consider one of them more robust than the other.

For external galaxies, we make use of 2D \textsc{galfit} \citep{2002AJ....124..266P,Peng2010} models obtained by \citetalias{Salo2015} in the framework of the S$^4$G project. We divide their single-disk models into two groups (see Table~\ref{tab:ref_table}): 1287 non-edge-on disk models and 217 edge-on models. We do so because edge-on and non-edge-on galaxies have different observed properties due to their inclination angle, such as the levels of PAH + dust emission and attenuation (although at 3.6~$\mu$m the last effect is drastically reduced), the visibility of different structural features (such as spiral arms, rings, bars) which can be easily revealed at specific inclination angles, and, of course, the resolution of the vertical structure which can only be studied for edge-on or close to edge-on galaxies. 

In addition, we also consider edge-on single-disk models from \citetalias{Bizyaev2009} and \citetalias{Mosenkov2010} that were obtained in the 2MASS \textit{K$_\mathrm{s}$} waveband.

\begin{figure}
    \centering
    \includegraphics[width=.5\textwidth]{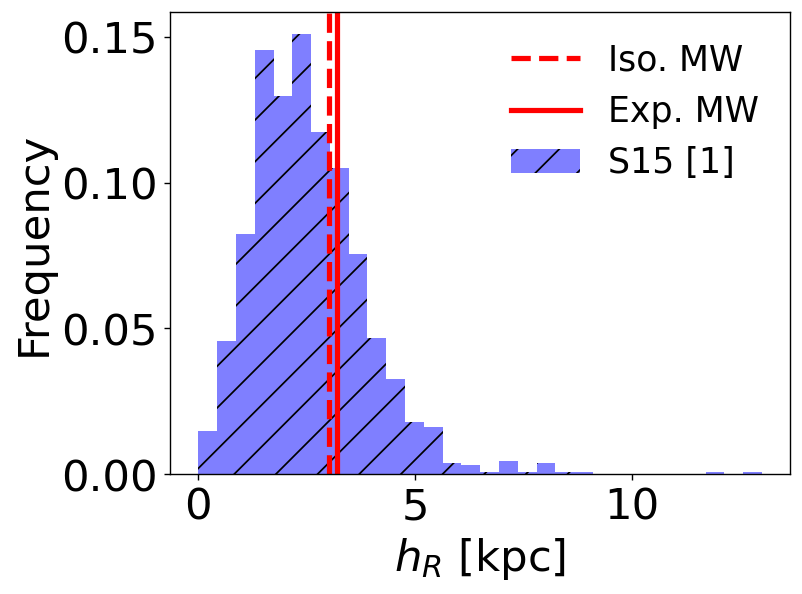}
    \caption{Distribution of the disk scale length for non-edge-on galaxies from \citetalias{Salo2015} [1] compared to the exponential (red solid line) and isothermal (red dashed line) Milky Way model scale lengths from \citetalias{Mosenkov2021}.}
    \label{fig:face_on_s4g}
\end{figure}

Figure~\ref{fig:face_on_s4g} demonstrates a distribution of the non-edge-on single-disk scale lengths for 1287 galaxy models from \citetalias{Salo2015}. Since these are non-edge-on galaxies, we have no data values for their scale heights, so we only show a histogram based on their scale lengths. We utilize the exponential and isothermal single-disk Milky Way model scale length values from \citetalias{Mosenkov2021} as a comparison.

\begin{figure*}
    \centering
    \includegraphics[width=.33\textwidth,height=3.33cm]{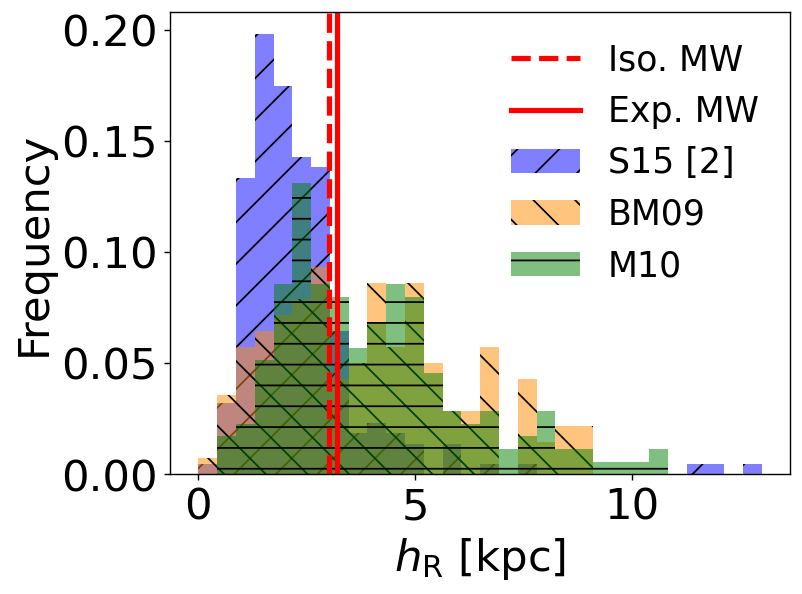}\hfill
    \includegraphics[width=.33\textwidth,height=3.35cm]{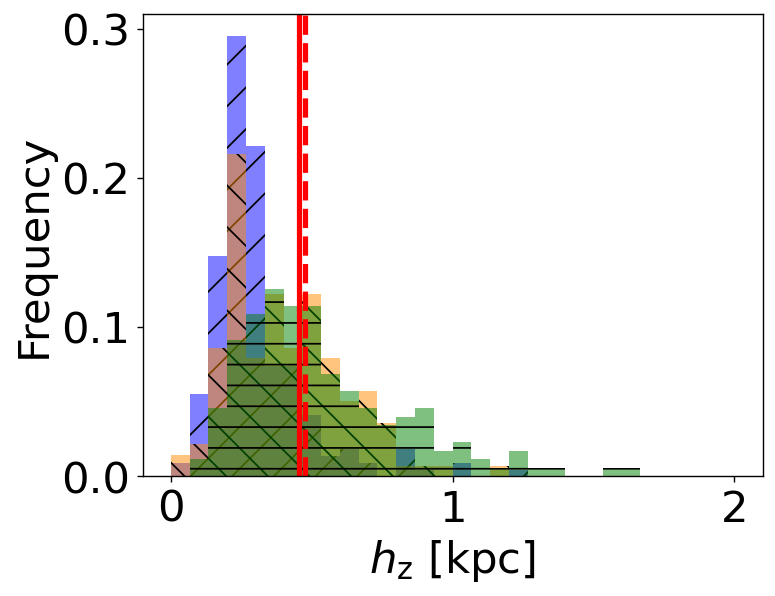}\hfill
    \includegraphics[width=.33\textwidth,height=3.43cm]{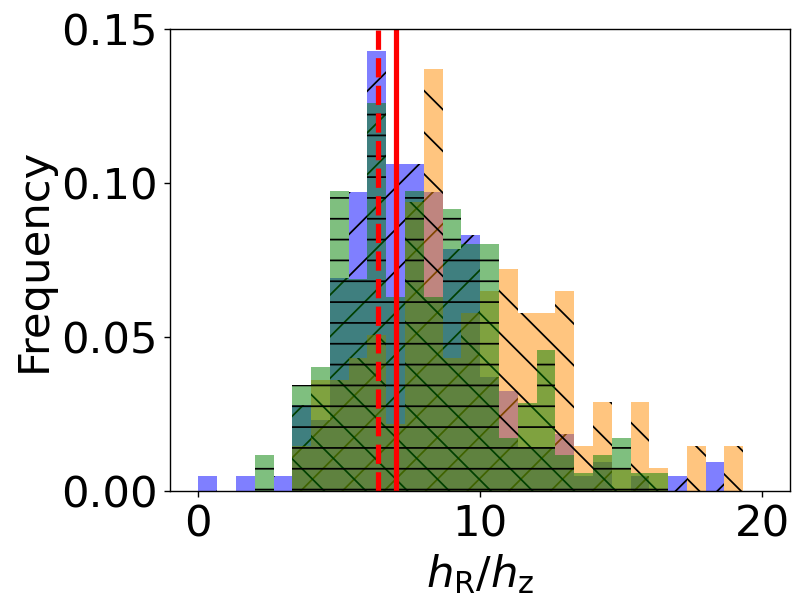}
    \\[\smallskipamount]
    \caption{Histograms of the scale length (\textit{left}), scale height (\textit{middle}), and ratio between the two (\textit{right}) for the NIR edge-on single-disk galaxy models from \citetalias{Bizyaev2009}, \citetalias{Mosenkov2010}, and \citetalias{Salo2015} [2]. The single-disk exponential and isothermal Milky Way model parameter values from \citetalias{Mosenkov2021} are highlighted by the solid and dashed red lines, respectively.}\label{fig:multi_single_graphs}
\end{figure*}

\begin{figure*}
    \centering
    \includegraphics[width=.48\textwidth]{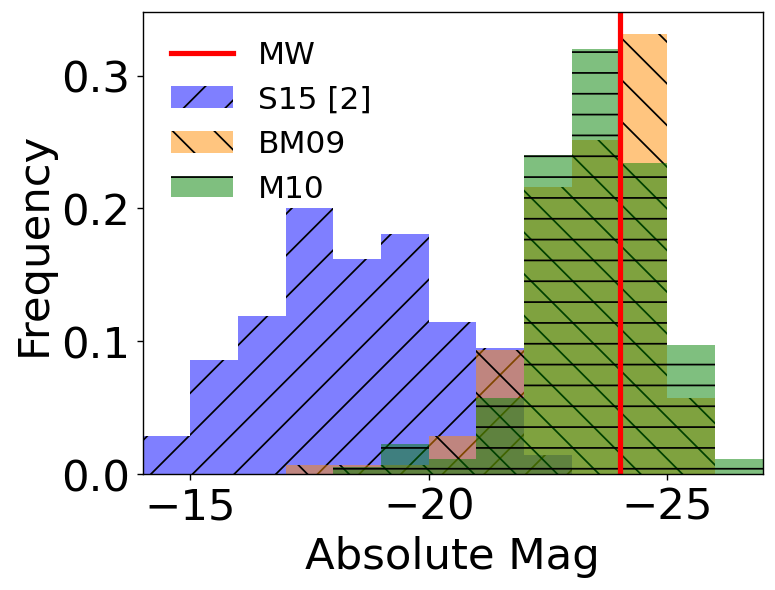}\hfill
    \includegraphics[width=.48\textwidth]{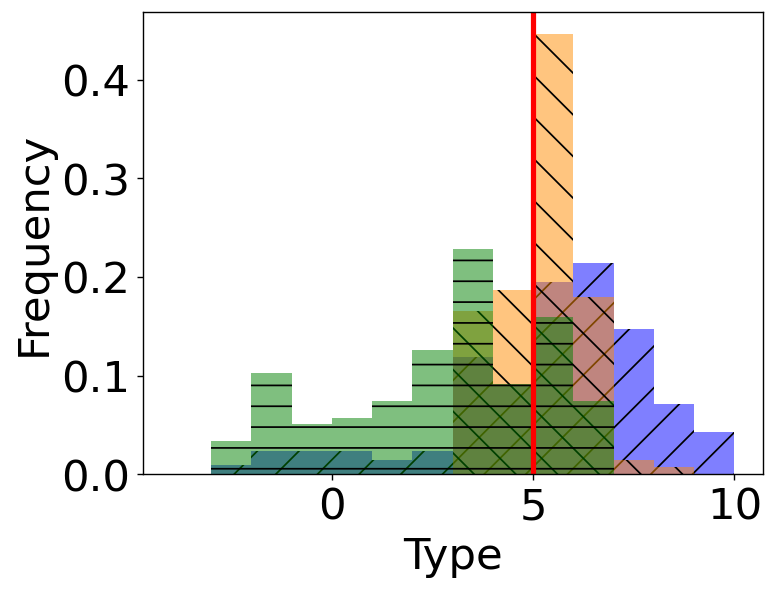}\hfill
    \\[\smallskipamount]
    \caption{Comparison of the absolute magnitudes (\textit{left}) and morphological types (\textit{right}) of galaxies from our NIR edge-on single-disk galaxy comparison (\citetalias{Bizyaev2009,Mosenkov2010,Salo2015} [2]). Milky Way values for both \textit{K}-band absolute magnitude \citep{Drimmel2001} and Type \citep{Georgelin1976} are highlighted.}\label{fig:multi_single_mag_morph}
\end{figure*}

Let us now turn to edge-on galaxies for which NIR single-disk models are available in the literature. Here we consider 139 galaxies from \citetalias{Bizyaev2009}, 175 galaxies from \citetalias{Mosenkov2010}, and 217 edge-on single-disk models from S$^4$G. The comparison of these datasets for the disk scale length, disk scale height, and the ratio of the two is displayed in Figure~\ref{fig:multi_single_graphs}. As one can see, the edge-on S$^4$G models appear to have lower values, on average, than the other two datasets for both scale length and scale height. The agreement between all \citetalias{Mosenkov2010} and \citetalias{Bizyaev2009} distributions is good. Although the S$^4$G models were obtained using {\it Spitzer} 3.6~$\mu$m observations while \citetalias{Mosenkov2010} and \citetalias{Bizyaev2009} used 2MASS \textit{K$_s$} data, the significant difference between the 2MASS and {\it Spitzer} data cannot be explained by the difference of the wavebands used. The \textit{K$_s$} (2.16~$\mu$m) and 3.6~$\mu$m wavebands are close and both map the old stellar population with little effects from the galaxy ISM. Therefore, the main reason for this discrepancy should be the morphological difference of the galaxies contained in the samples. To confirm this, we present Figure~\ref{fig:multi_single_mag_morph}, which compares the absolute magnitudes and morphological types of the samples. We note here that while the classification of morphologies of edge-on galaxies is more subjective than that of face-on galaxies, there are methods that can be used to determine these morphologies, such as calculating the contribution of the bulge mass to the total mass \citep{Obreschkow2009}. For the galaxies in Figure~\ref{fig:multi_single_mag_morph}, the morphology data comes from HyperLeda\footnote{http://leda.univ-lyon1.fr/}. As one can see, the edge-on galaxies from \citetalias{Salo2015} host systematically less luminous stellar disks and are of later morphological types as compared to the \citetalias{Bizyaev2009} and \citetalias{Mosenkov2010} samples. Due to this bias within the \citetalias{Salo2015} sample, we do not include it within our calculation of the quartiles for this comparison.

Using the histograms presented in this section, we can draw the following conclusions. Beginning with the non-edge-on galaxies in Figure~\ref{fig:face_on_s4g}, it is clear that both values of the Milky Way's scale length (the exponential and isothermal model values) are slightly larger than average, but not by a substantial amount. This is proven by the median and Q3 cutoff values for this dataset, showing that both Milky Way model scale length values are only slightly higher than the median. 

Moving on to the edge-on single-disk models in the NIR (see Fig.~\ref{fig:multi_single_graphs}), we see more of the pattern of normality previously established by the non-edge-on single-disk NIR galaxy models. Both scale length values are slightly below the median but above the Q1 cutoff value. Thus, it is evident that the scale length values are typical of spiral galaxies. Similarly, both scale height values fall just above the median value. 

The two ratios of the scale length and scale height values for the Milky Way in Figure~\ref{fig:multi_single_graphs} are largely unremarkable. The exponential model value is slightly larger than the isothermal model value, although both are close to Q2 and, as such, can be considered normal. 

\begin{figure*}
    \centering
    \includegraphics[width=.57\textwidth,height=6.9cm]{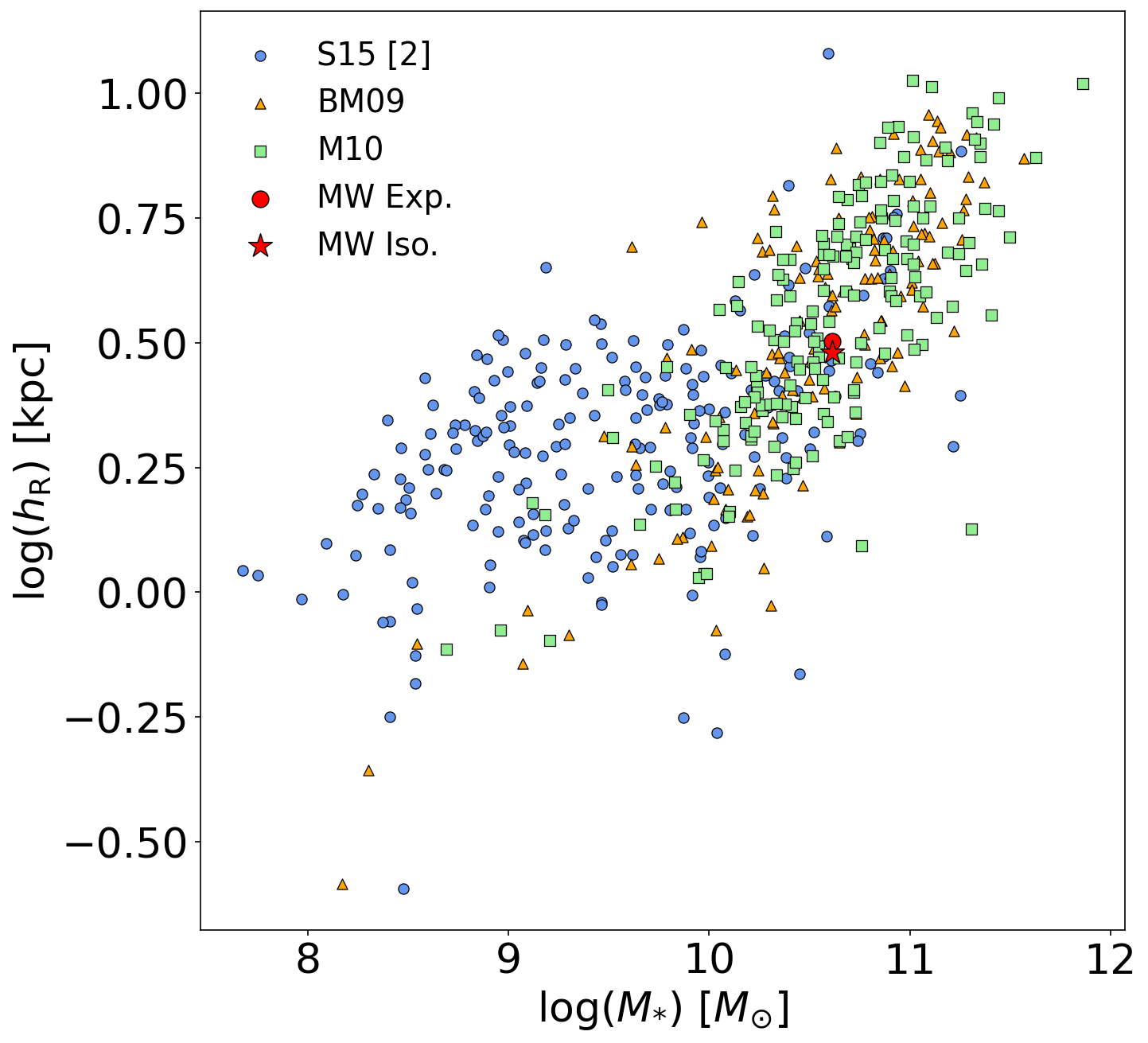}\hfill
    \\[\smallskipamount]
    \caption{Stellar mass versus scale length for our NIR edge-on single-disk data samples (\citetalias{Salo2015} [2]; \citetalias{Bizyaev2009}; \citetalias{Mosenkov2010}), though the stellar masses for \citetalias{Salo2015} [2] come from \citetalias{Sheth2010}. The Milky Way mass value comes from \citet{Bland2016}, while the exponential (\textit{red circle}) and isothermal (\textit{red star}) scale length values come from \citetalias{Mosenkov2021}.}
    \label{fig:NIR_hr_mass}
\end{figure*}

We can thus conclude that in NIR single-disk decompositions, the Milky Way's structural parameters are all well within the range of typical values. We also demonstrate this in Figure~\ref{fig:NIR_hr_mass} -- the Milky Way perfectly obeys the correlation between the disk scale length and stellar mass.

\section{Two-disk NIR Waveband Galaxy Model}
\label{sec:doubleIR}

\begin{figure*}
    \centering
    \includegraphics[width=.48\textwidth]{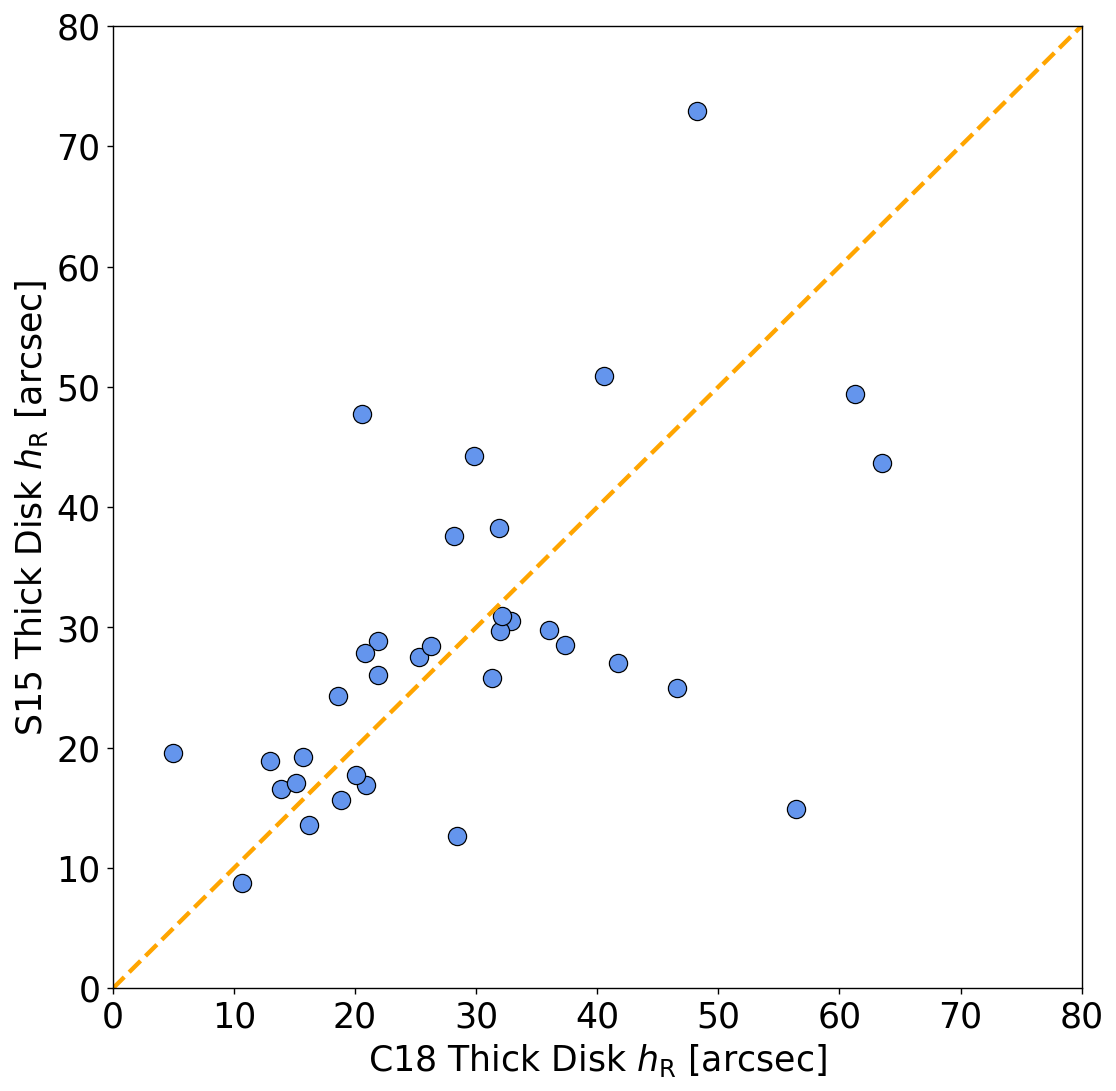}\hfill
    \includegraphics[width=.50\textwidth,height=6.8cm]{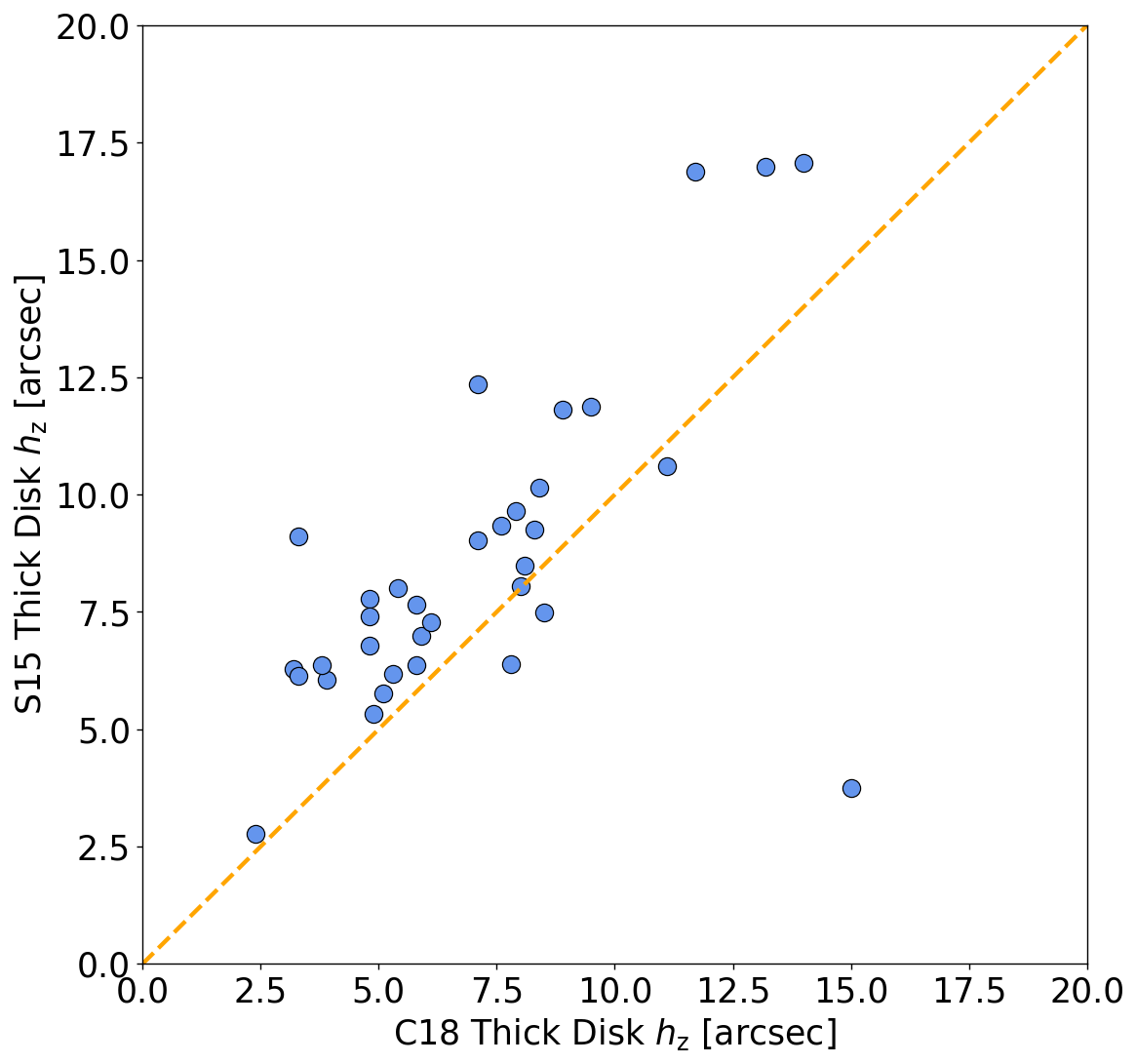}\hfill
    \includegraphics[width=.48\textwidth,height=6.95cm]{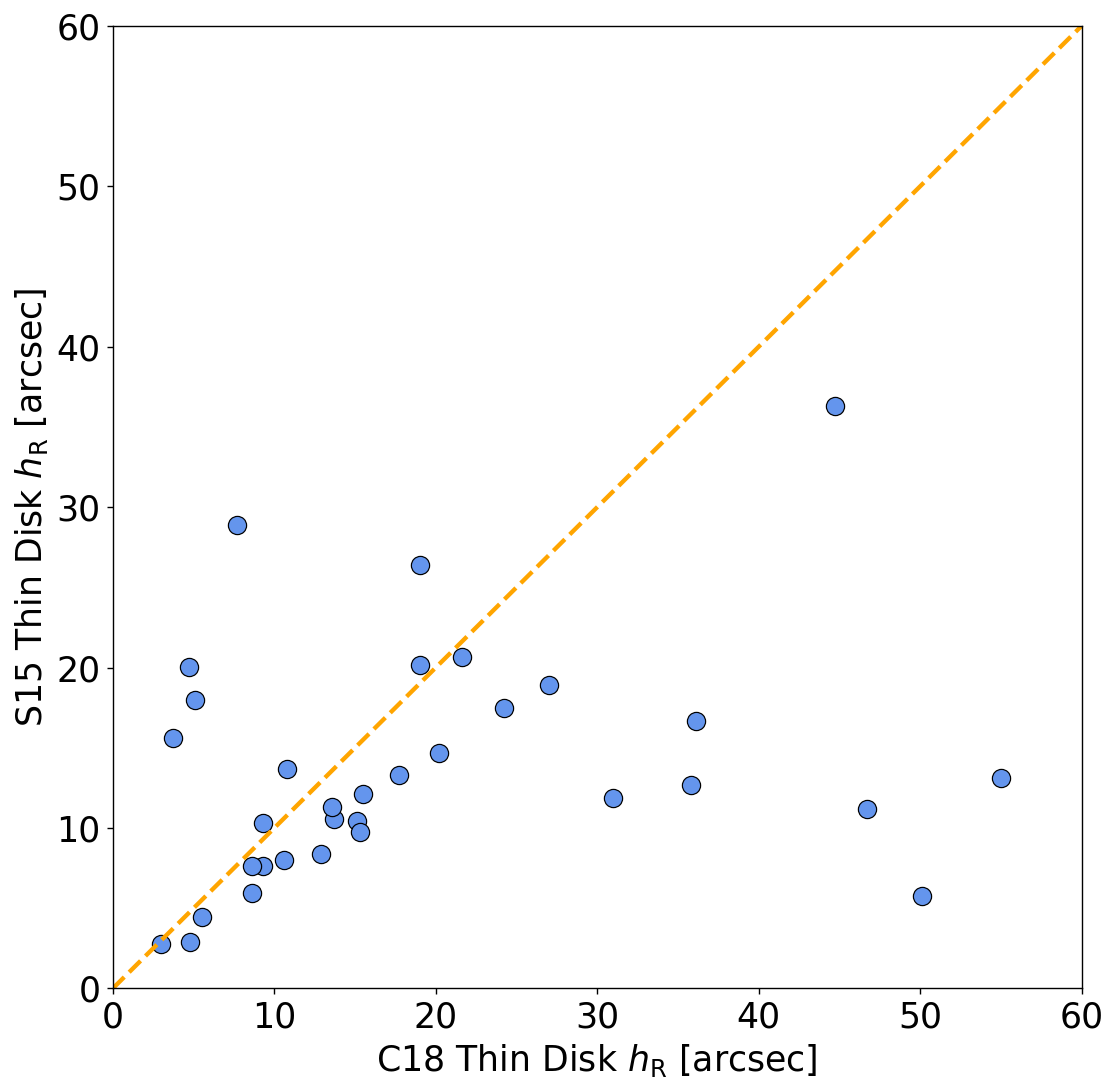}\hfill
    \includegraphics[width=.48\textwidth]{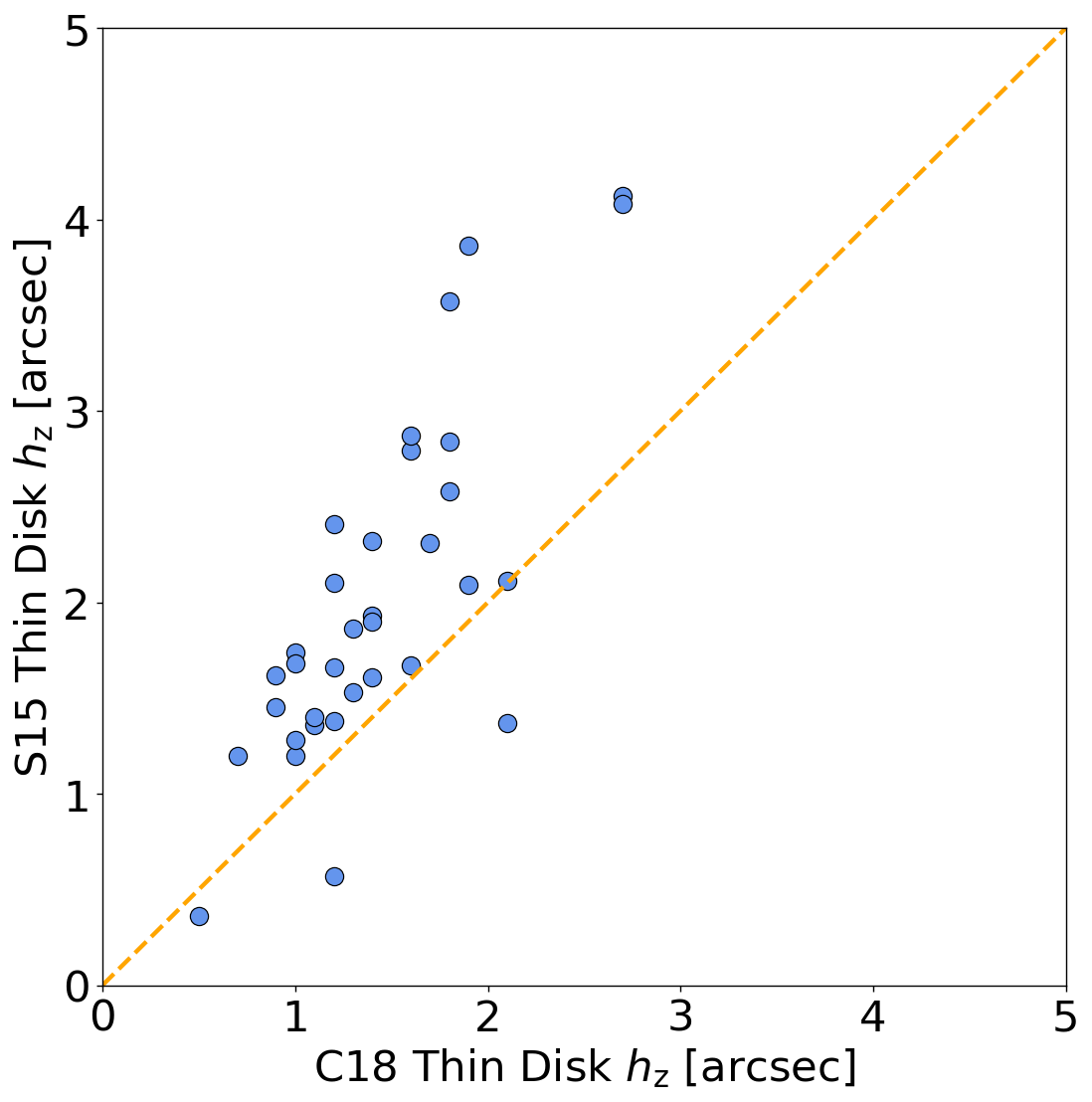}\hfill
    \\[\smallskipamount]
    \caption{Scale length and scale height comparisons of the same 34 galaxies from \citetalias{Salo2015} [3] and \citetalias{Comeron2018} [1]. A dashed orange line representing $y=x$ is included to aid in the comparison.}\label{fig:salo_comeron_scale_comparison}
\end{figure*}

We now move on to the two-disk scale length and scale height comparisons. For these comparisons, we utilize edge-on two-disk decomposition models created in the NIR waveband, the only spectral region where we have a significant number of galaxy models consisting of two photometrically-decoupled stellar disks. The difference between the surface brightness profiles of the thin and thick disks is that the scale height of the thick disk is significantly larger than the scale height of the thin disk in a photometric rather than chemical context \citep{Burstein1979,Walker1996,Schonrich2009,Bornaud2009,Beraldo2021}. Such a two-disk model of the Milky Way is also supported by a double exponential vertical {\it number density} profile \citep{Yoshii1982,Gilmore1983,Chen2001,Ojha2001,Chrobakova2020}. There are other distinctive features of the thin and thick disks such as the chemical composition and kinematics, but the surface brightness profile of each is of the most interest to us in this study. Within \citetalias{Mosenkov2021}, this two-disk decomposition model best fits the observed data of the Milky Way, with the exponential model being slightly more accurate than the isothermal model (see Sect.~\ref{sec:introduction}). As such, the analysis of the scale length and scale height in this section will be of the most worth for exploring the formation scenario of the thin and thick disks in our Galaxy in comparison with other spiral galaxies.

As with the single-disk models, we begin by gathering both the exponential and isothermal models of the Milky Way from \citetalias{Mosenkov2021}. These models also contain outer flaring, warping, and two blobs on either side of the Galactic center representing most apparent over-densities. In comparison to the other Milky Way models we have used, the disk scale height and disk flatness values for these two models are more different from each other, which provides for some interesting situations in our analysis where one model's value appears abnormal and the other appears average.

For external galaxies, we gather our comparison data from four different sources: \citetalias{Salo2015}, \citetalias{Comeron2018}, \citetalias{Mosenkov2018}, and \citetalias{Pinna2023}. We note here that \citetalias{Pinna2023} only provide scale height data and gather their data using optical wavebands, so the comparison to their values is not entirely justified. However, we include them in order to show the large discrepancy between the Milky Way and other Milky Way-mass galaxies, so while the details might not be accurate, the overall comparison still provides important information. As with the 2MASS data from the edge-on single-disk NIR galaxies, both \citetalias{Salo2015} and \citetalias{Comeron2018} provide models of S$^4$G galaxies, but because they utilize different fitting methods, their values are different and thus we consider both datasets separately. Specifically, \citetalias{Comeron2018} utilize a more extended point spread function (PSF) than \citetalias{Salo2015} which likely causes the thin and thick disk scale heights of galaxies contained in both datasets to be systematically smaller in the \citetalias{Comeron2018} models (see Fig.~\ref{fig:salo_comeron_scale_comparison} where we compare common galaxies in both samples). The influence of the PSF on measured galaxy observables was studied in detail in \citet{2014A&A...567A..97S} and \citet{2015A&A...577A.106S} which indeed suggest that the vertical structure of edge-on galaxies may be significantly affected by the instrumental scattered light. Also, the difference in the results may be in the different methods used in these studies.  However, the \textit{full} \citetalias{Comeron2018} dataset demonstrates higher scale heights, on average, due to \citetalias{Salo2015} being biased towards dimmer galaxies (similar to the \textit{left} panel in Fig.~\ref{fig:multi_single_mag_morph}).

As part of their decomposition, \citetalias{Comeron2018} inspected the axial surface-brightness profiles of the S$^4$G edge-on galaxies in order to choose how many exponential segments each disk contained. They were able to accomplish this because the edge-on line-of-sight profiles of the galaxies were similar to piecewise exponential profiles and thus could be separated into different segments, each with its own scale length (see their sections 3.3.2 and 3.3.3 for further details on the functions used and the fittings performed). Since the exponential and isothermal Milky Way models created by \citetalias{Mosenkov2021} only have one scale length for each of the thin and thick disks (the Galaxy radial surface brightness profile does not demonstrate a significant brake or truncation), we use the radial scale length from the first (closest to the center) exponential section of each galaxy fitted by \citetalias{Comeron2018}.

\begin{figure*}[!t]
    \centering
    \includegraphics[width=.32\textwidth,height=3.33cm]{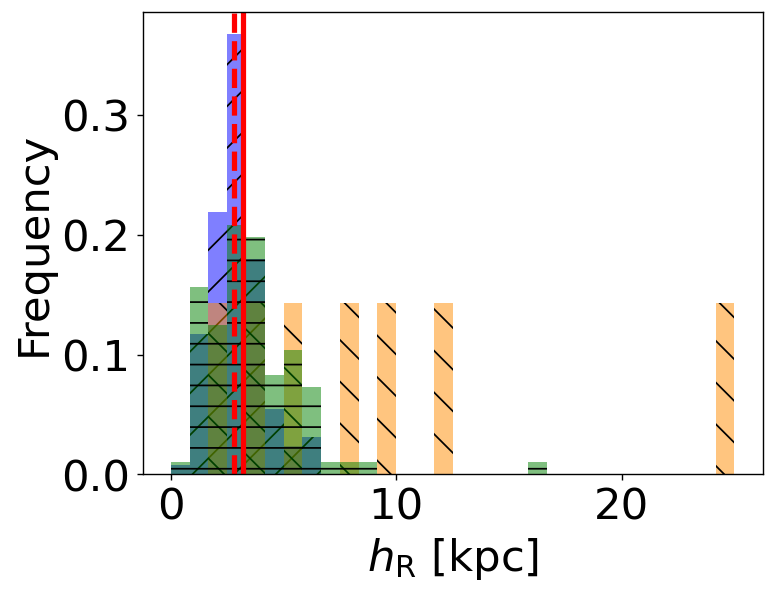}\hfill
    \includegraphics[width=.32\textwidth,height=3.33cm]{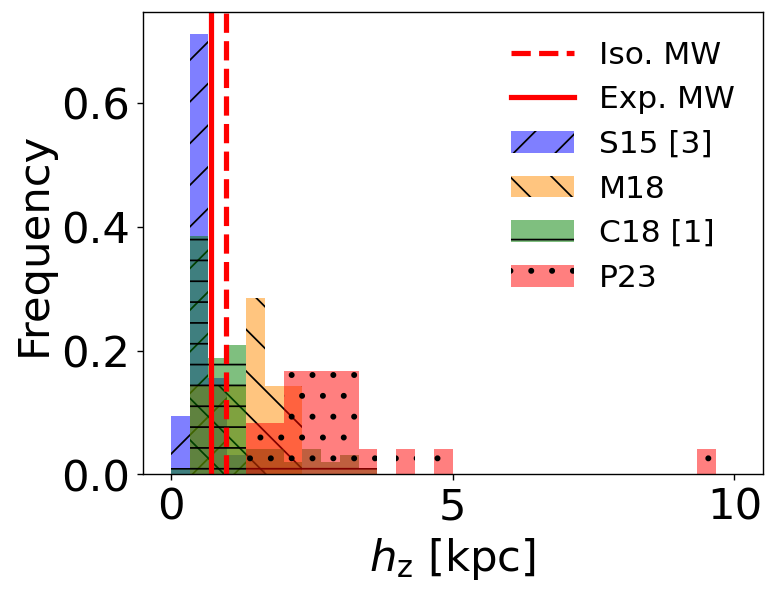}\hfill
    \includegraphics[width=.32\textwidth,height=3.44cm]{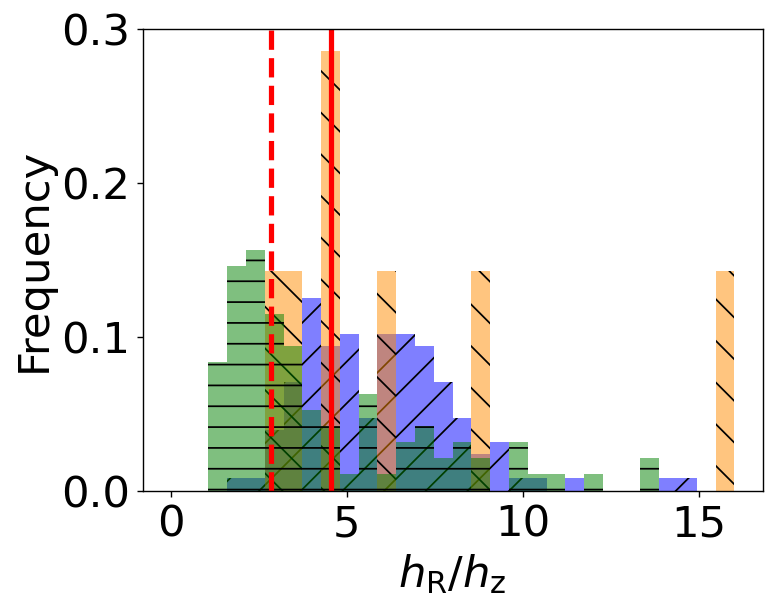}
    \\[\smallskipamount]
    \includegraphics[width=.32\textwidth,height=3.33cm]{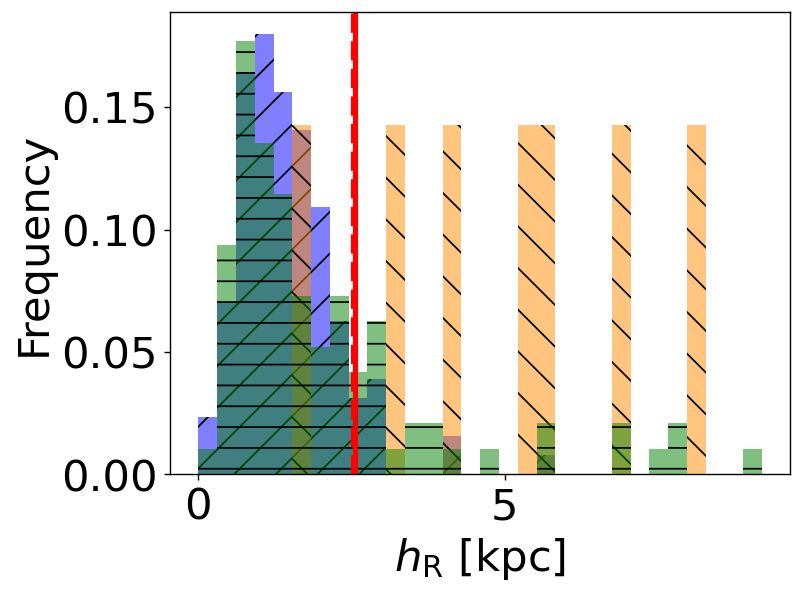}\hfill
    \includegraphics[width=.32\textwidth,height=3.33cm]{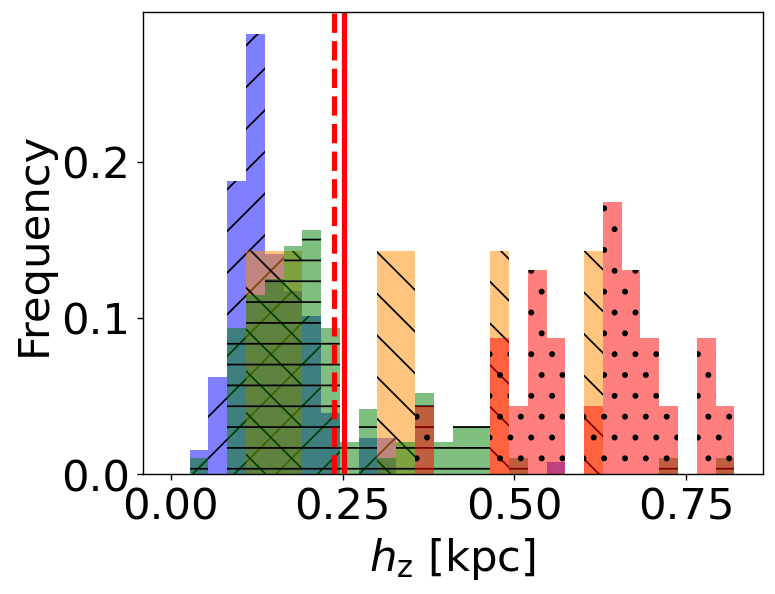}\hfill
    \includegraphics[width=.32\textwidth,height=3.44cm]{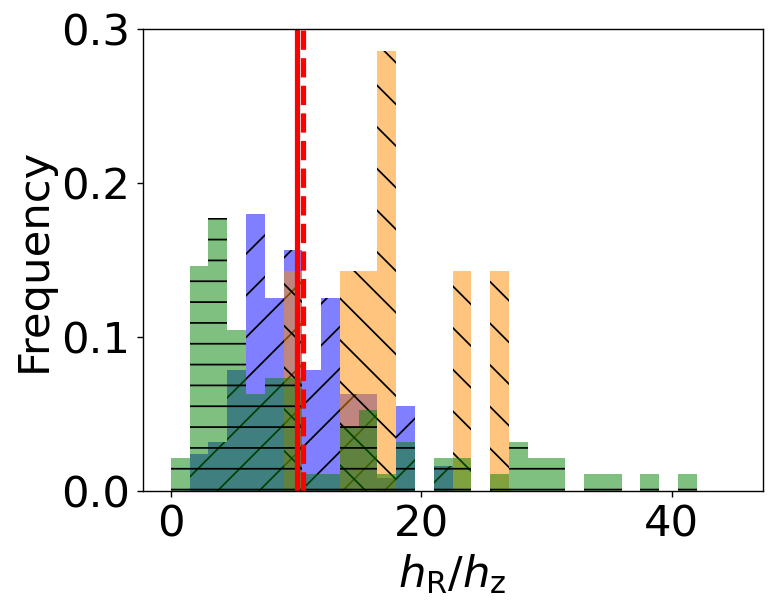}
    \caption{NIR waveband edge-on two-disk galaxy models (\citetalias{Salo2015} [3]; \citetalias{Mosenkov2018,Comeron2018} [1]; \citetalias{Pinna2023}). As a note, \citetalias{Pinna2023} use optical wavebands; see text for the reasoning of their inclusion in this plot. Histograms for both disks are shown. \textit{Top Row}: Thick disk galaxy data. Histograms of the scale length (\textit{left}, in kpc), scale height (\textit{middle}, in kpc), and ratio between the two (\textit{right}) are shown. The thick disk exponential and isothermal Milky Way model parameter values from \citetalias{Mosenkov2021} are highlighted in each plot by the solid red line and the dashed red line, respectively. \textit{Bottom Row}: Thin disk galaxy data. The order of the histograms is the same as the top row, with the respective thin disk exponential and isothermal Milky Way model parameter values from \citetalias{Mosenkov2021} highlighted.}\label{fig:multi_2disk_graphs}
\end{figure*}

Using the combined data presented in Figure~\ref{fig:multi_2disk_graphs}, we can compare the Milky Way's two-disk model to external galaxies. From an initial look at the thick disk models, it \textit{appears} that the Milky Way has relatively standard values. For the thick disk scale length, scale height, and flatness ratio, the Milky Way's values are mostly within the second or third quarters for each parameter (with the exception being the isothermal flatness ratio, which falls within the first quarter). \textit{However}, it is important to point out that our total data sample is dominated by S$^4$G galaxies, which we have shown to be of sytematically lower mass (see Figs.~\ref{fig:optical_Ropt_mass}, \ref{fig:multi_single_mag_morph}, and \ref{fig:NIR_hr_mass}). The Milky Way-mass galaxies from \citetalias{Mosenkov2018} and \citetalias{Pinna2023} (from the AURIGA zoom-in cosmological simulation) indicate that the Milky Way's thick disk is significantly thinner (smaller scale height value) and much less extended (smaller scale length value) than would be expected of a galaxy of its stellar mass. Thus, while at first glance the Milky Way's thick disk might appear normal, it is actually only normal for galaxies of a much smaller mass, indicating it is far smaller than one would anticipate.

Moving on to the thin disk data (the bottom row in Fig.~\ref{fig:multi_2disk_graphs}), we see that the scale length and scale height values of our Galaxy fall on the upper end of the distribution of S$^4$G galaxies, though, like the thick disk, both are below the values of the galaxies from \citetalias{Mosenkov2018} and \citetalias{Pinna2023}. At first glance, this would seem to imply that the Milky Way has a relatively large thin disk; however, keeping in mind the same points from the previous paragraph, this actually tells us that, while the Milky Way's thin disk is larger than lower-mass galaxies, it is still thinner and less extended than other Milky Way-mass galaxies. While less abnormal than the thick disk, the Galaxy's thin disk still appears slightly smaller compared to the \textit{HER}OES and AURIGA galaxies with similar stellar masses.

\subsection{\texorpdfstring{S\textsuperscript{4}G}{S4G} Disk Mass Data and Analysis}
\label{subsec:disk_mass}

\begin{figure*}[t!]
    \centering
    \includegraphics[width=.5\textwidth]{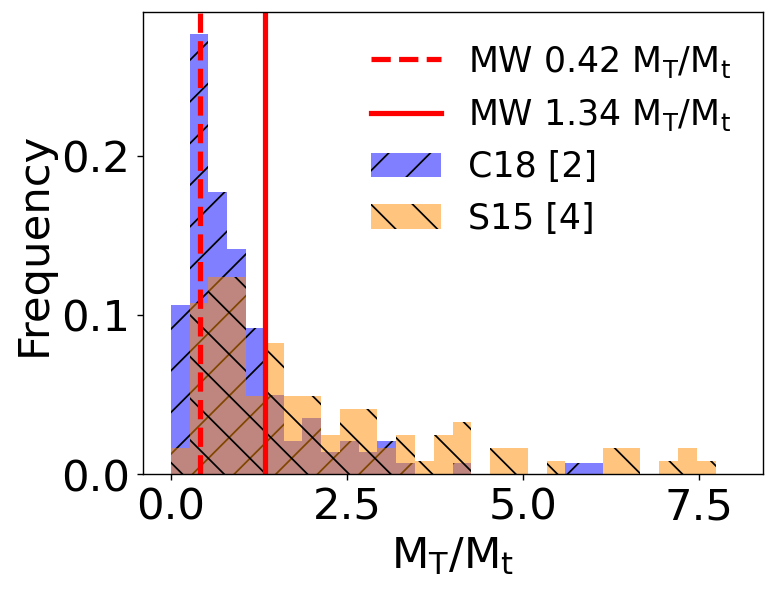}
    \caption{Disk mass ratios of galaxies from \citetalias{Comeron2018} [2] and \citetalias{Salo2015} [4]. The minimum and maximum values of the Milky Way's disk mass ratio from \citetalias{Mosenkov2021} are highlighted. \label{fig:Mass_ratio_graphs}}\hfill
    \\[\smallskipamount]
\end{figure*}

\begin{figure*}[t!]
    \includegraphics[width=0.48\textwidth]{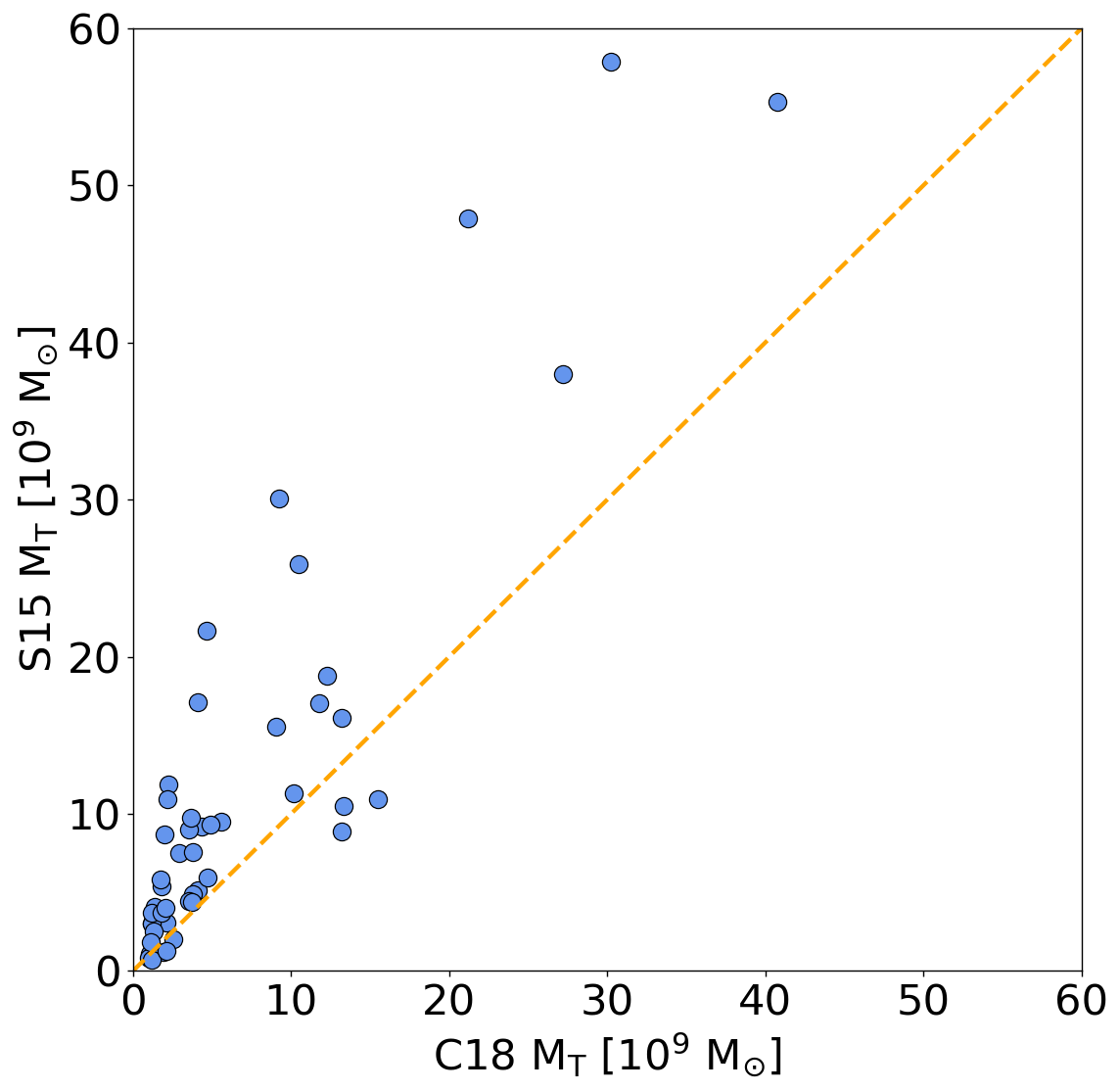}\hfill
    \includegraphics[width=0.48\textwidth]{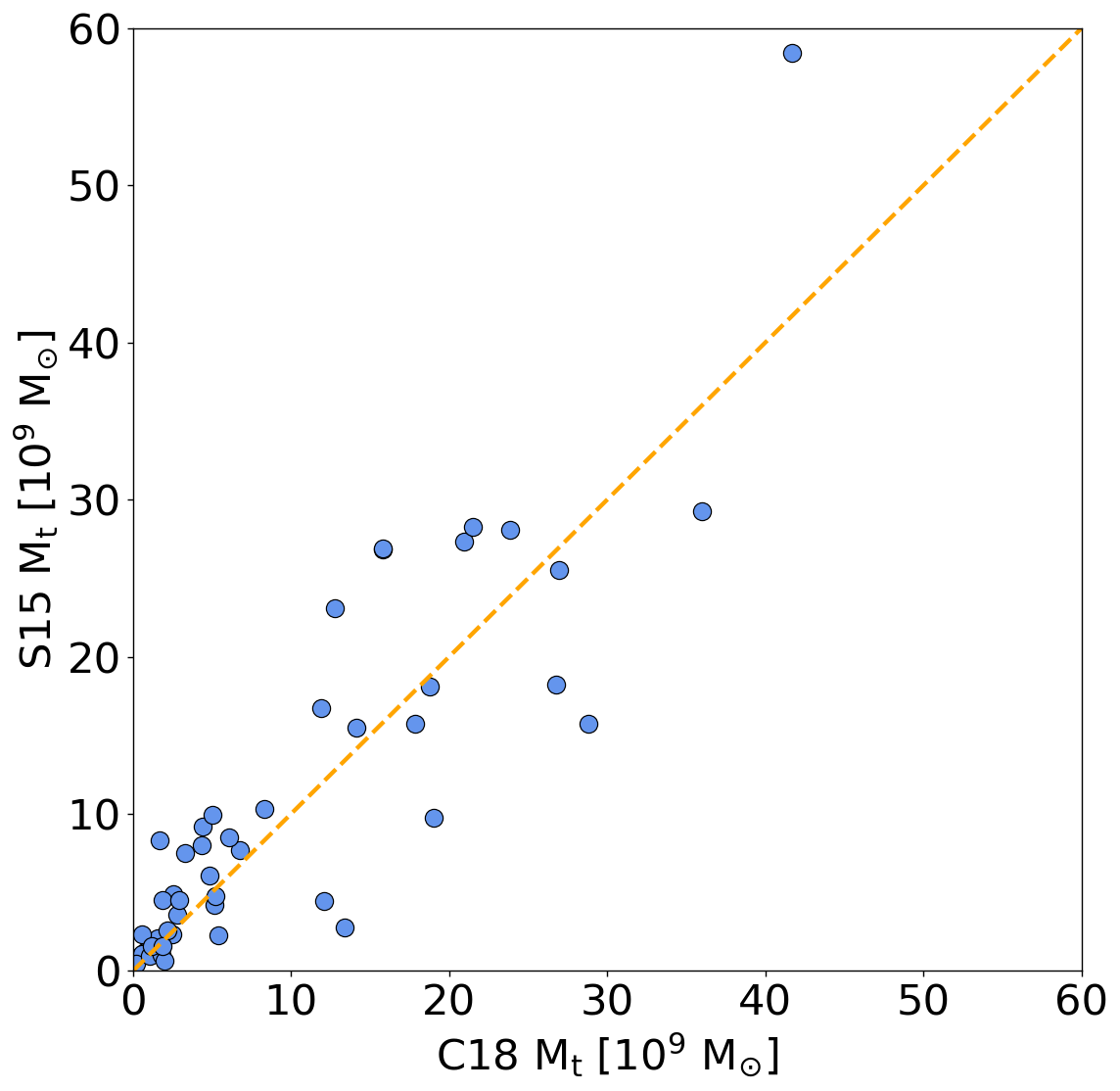}\hfill
    \\[\smallskipamount]
    \caption{A comparison of the disk masses (in 10$^9$ solar masses) of the same 48 galaxies from \citetalias{Comeron2018} [2] and \citetalias{Salo2015} [4]. Both the thick disk (\textit{left}) and thin disk (\textit{right}) data are shown. On each plot, the line $y=x$ is shown to help better visualize the difference between the two paper samples.
    \label{fig:Mass_difference_graphs}}
\end{figure*}

\begin{figure*}[t!]
    \centering
    \includegraphics[width=0.55\textwidth]{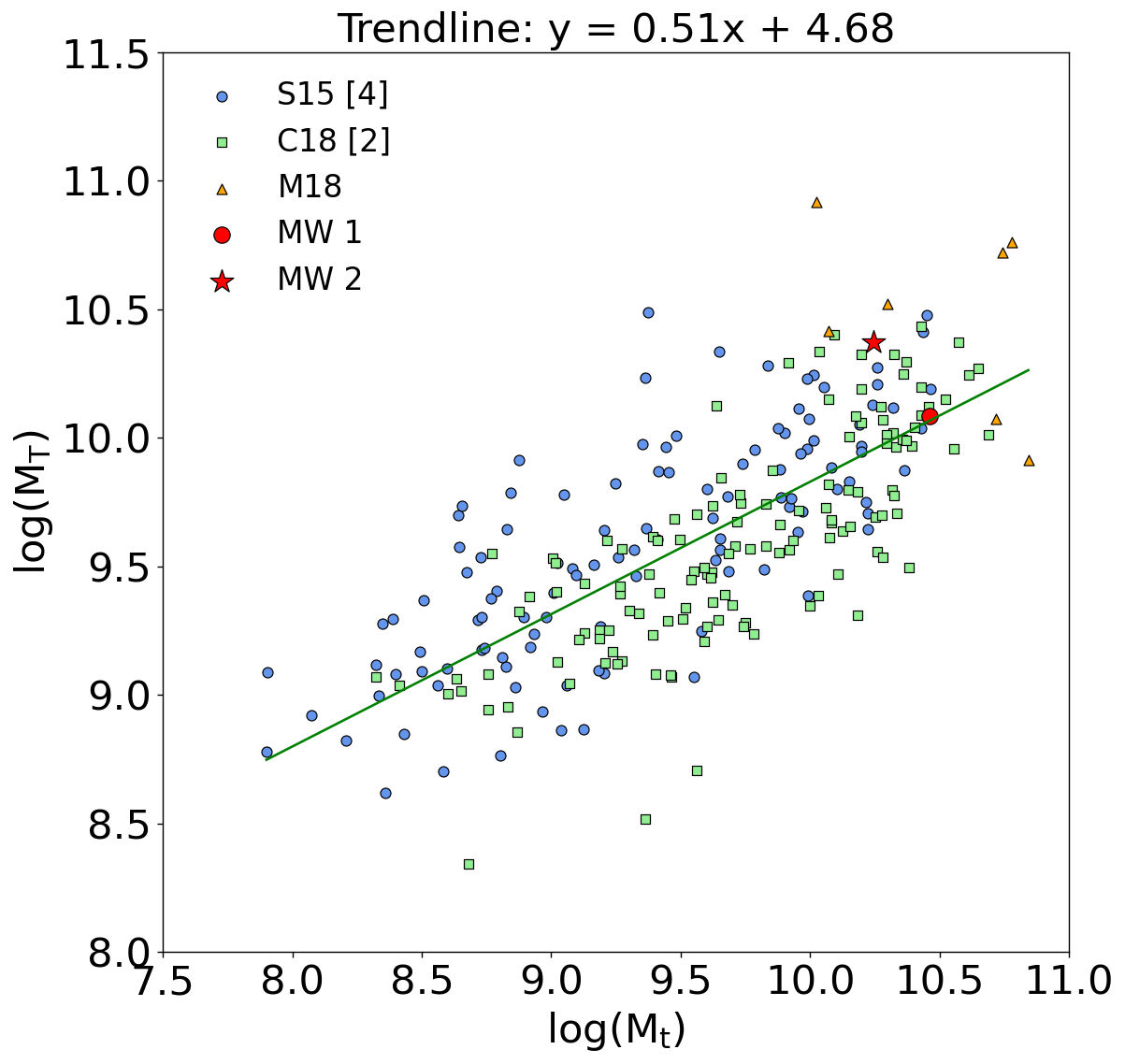}\hfill
    \\[\smallskipamount]
    \caption{Thin and thick disk masses for various galaxies from \citetalias{Salo2015} [4], \citetalias{Comeron2018} [2], and \citetalias{Mosenkov2018}. The masses are given in 10$^9$ solar masses. A trendline (\textit{green}) is included in the plot to show the trend of the distribution of data (with the trendline equation being displayed above the  plot). The Pearson correlation coefficient for the trendline is 0.74. The two Milky Way datapoints included represent the 0.42 disk mass ratio (red dot) and the 1.34 disk mass ratio (red star).}\label{fig:Mass_scatter_graphs}
\end{figure*}

Within this section we discuss models we gathered concerning the disk masses of the S$^4$G galaxies from \citetalias{Salo2015} and \citetalias{Comeron2018}. We analyze the thick-disk-to-thin-disk mass ratio (see Fig.~\ref{fig:Mass_ratio_graphs}), the difference between the same galaxy's mass values in \citetalias{Salo2015} and \citetalias{Comeron2018} (see Fig.~\ref{fig:Mass_difference_graphs}), and the thick and thin disk masses of each galaxy (see Fig.~\ref{fig:Mass_scatter_graphs}). As before, for the Milky Way we use the models from \citetalias{Mosenkov2021}. However, the only data they provide is the disk mass ratio, so we utilize a total disk mass value of $4.1\times10^{10}$~$M_{\sun}$ from \citet{Bland2016} and use the two ratio values to calculate individual thick and thin disk mass values (details on why there are two values are provided later).

We begin by describing the process through which we obtained the stellar masses of the disk models. \citetalias{Comeron2018} provide the masses of the thick and thin disks, so we use their masses as they are. In \citetalias{Salo2015}, however, the stellar masses were not computed; instead, there was data for each disk concerning the scale length, scale height, edge-on central surface brightness, and distance (the galaxy distances were originally provided in \citealt{Sheth2010}), which was enough to allow us to calculate the disk masses. To compute the apparent face-on central surface brightness of an isothermal disk
\begin{equation}
    I(r,z)=I(0,0)\,\mathrm{e}^{-r/h_\mathrm{R}}\mathrm{sech}^2(z/z_0)\,,
\end{equation}
we need to integrate its profile along the $z$ axis at $r=0$, so
\begin{equation}
    I_0^{\mathrm{face-on}}=I(0,0)\int_{-\infty}^{\infty}\mathrm{sech}^2(z/z_0)\mathrm{dz}=2\,z_0I(0,0)\,.
\end{equation}
In a similar way, we can show that the apparent edge-on central surface brightness is
\begin{equation}
    I_0^{\mathrm{edge-on}}=2\,h_{\mathrm{R}}I(0,0)\,.
\end{equation}
Hence, we can find the relationship between the apparent edge-on and face-on central surface brightness through
\begin{equation}
    I_0^{\mathrm{edge-on}}=I_0^{\mathrm{face-on}}h_{\mathrm{R}}/z_0\,.
\end{equation}
Finally, we arrive at
\begin{equation}
     \mu_0^{\mathrm{face-on}} =  \mu_0^{\mathrm{edge-on}} + 2.5\,\log{h_{\mathrm{R}}/z_0} \,.
\end{equation}
Similarly, it can be shown that for an exponential disk, the face-on central surface brightness is related to the edge-on central surface brightness via
\begin{equation}
     \mu_0^{\mathrm{face-on}} = \mu_0^{\mathrm{edge-on}} + 2.5\,\log{h_{\mathrm{R}}/h_{\mathrm{z}}} \,.
\end{equation}

Then, using the newly-calculated face-on central surface brightness and the scale length, we calculated the total apparent magnitude of each disk using
\begin{equation}
    m_{\mathrm{tot}}=\mu_0^{\mathrm{face-on}}-5\,\log(h_\mathrm{R})-1.955,
\end{equation}
where $m_{\mathrm{tot}}$ is the total apparent magnitude. Once we had calculated the apparent magnitude of each disk, we used the distances to calculate the absolute magnitudes of the galaxies. Finally, we used the following equation from the literature (\citealt{MunozMateos2013,DiazGarcia2016}; see also \citealt{Eskew2012}) to calculate the mass of each disk in solar masses:
\begin{equation}
    \log(M_*/M_{\odot})=-0.4\,M_{3.6\mathrm{{\mu}m}}+2.13
\end{equation}
where $M_*$ is the disk's mass and $M_{3.6\mathrm{{\mu}m}}$ is the absolute magnitude at 3.6~$\mu$m.

In Figure~\ref{fig:Mass_ratio_graphs}, we display a histogram showing the ratio of the thick disk mass over the thin disk mass. By observation, we can see that the data from \citetalias{Comeron2018} (denoted by the blue bars in the histogram) contains slightly lower values, on average, than the data from \citetalias{Salo2015} (denoted by the orange bars). The Milky Way data from \citetalias{Mosenkov2021}, displayed in this histogram, contains two disk mass ratio values, representing the minimum and maximum ratio values. The difference between these minimum and maximum values is considerable, but even the minimum $M_{\mathrm{T}}/M_{\mathrm{t}} = 0.42$ implies that the contribution of the thick disk to the Galaxy mass budget is substantial. 

This range of values is due to the uncertainty in the local thick-to-thin disk surface density ratio. Specifically, two values are given: $f_{\Sigma} = 0.38$ from \citet{Snaith2014} and $f_{\Sigma} = 0.12$ from \citet{Bland2016}. The reason for this uncertainty is because the local density normalization of the Milky Way (the ratio of the thick disk density over the thin disk density) is correlated to the scale height of the thick disk (higher scale height estimates result in lower normalization estimates and vice versa, \citealt{Bland2016}). Since the scale height values for both Milky Way disks have a range of estimated values in the literature, the normalization is poorly constrained, and thus the local surface density ratio is also poorly constrained since they are proportional to each other.

The Milky Way's low-end disk mass ratio value falls in the first quarter of the distribution in Figure~\ref{fig:Mass_ratio_graphs}, while the high-end value lies amidst some of the mid-sized values of our data samples within the third quarter. We see from this comparison that the low Milky Way value, being smaller (though not an outlier) than the average value for other galaxies, is still significant and much higher than usually estimated with other methods (\citealt{Bland2016} give $M_\mathrm{T}/M_\mathrm{t}=0.17$ which lies well below Q1). Therefore, our estimate of $M_\mathrm{T}/M_\mathrm{t}$ is more realistic and signifies that the contribution of the thick disk to the total mass budget in spiral galaxies is substantial, if not dominant.

Figure~\ref{fig:Mass_difference_graphs} shows the difference in disk mass values for the common 48 galaxies in \citetalias{Comeron2018} and \citetalias{Salo2015}. For the thin disk, the \citetalias{Salo2015} values are systematically higher than those from \citetalias{Comeron2018}, whereas for the thick disk, the general trend follows a one-to-one relation. This difference can be explained by the different PSFs and methods used in these studies. 

Figure~\ref{fig:Mass_scatter_graphs} displays a scatter plot of the thin disk's and thick disk's stellar masses from \citetalias{Salo2015} [4], \citetalias{Comeron2018} [2], and \citetalias{Mosenkov2018}. The Milky Way data points we use in these scatter plots come from a combination of sources: the total disk mass value comes from \citet{Bland2016} while the individual thick and thin disk mass values come from using the ratios discussed above \citepalias{Mosenkov2021} combined with the total disk mass value. Since we have two ratio values for the disk masses, we added two Milky Way data points on the graph. A trendline is included, with the equation being shown above the scatter plot. The Pearson correlation coefficient for the trendline is 0.74, indicating a relatively strong positive correlation.
In the scatter plot, the 0.42 and 1.34 disk-mass-ratio data points for the Milky Way lie relatively close the trendline (though we note that the 0.42 data point lies almost directly on the line), making our Galaxy quite typical in this context as well. We stress here that while the Milky Way is certainly more massive than the majority of the S$^4$G galaxies, the correlation between mass and size for spiral galaxies is not nearly as significant as it is for elliptical galaxies \citep{2014MNRAS.444..682C}, which suggests that normal spiral galaxies are not so dramatically homologously different. As such, the relationships considered in this section and the scale length and height comparisons performed in previous sections are still justified despite the difference in the stellar masses of the spiral galaxies under study.

\section{Bulge}
\label{sec:bulge}

We now analyze the bulge-to-total ($B/T$) flux ratio of the Milky Way against other galaxies. \citetalias{Mosenkov2021} detail the morphology of the Milky Way's bulge, confirming the boxy/peanut/X-shape previously described in the literature \citep{Dwek1995,McWilliam2010,Saito2011} and conclude through an \textit{N}-body model of the bar that the asymmetry in the Galaxy bulge/bar's shape is likely caused by our viewing angle of the bulge/bar (see fig. 6 in their paper). We utilize their value of the Milky Way's $B/T$ ratio as a maximum value for our analysis since they don't consider the bulge and bar separately, thus inflating the actual $B/T$ ratio with flux from the bar (technically, \citetalias{Mosenkov2021} give the bulge+bar-to-disk flux ratio, so we assumed a general decomposition of only a disk and a bulge+bar and then calculated the $B/T$ ratio using the bulge+bar-to-disk ratio). In an attempt to give a more accurate estimate of our Galaxy's $B/T$ ratio, we utilize a stellar mass ratio of the bulge and bar from \citet{Portail2017} to account for and remove the excess bar flux from the ratio, though this is just an estimate since, in principle, the mass-to-light ratio of the bulge and bar can be different.

\begin{figure*}
    \centering
    \includegraphics[width=.49\textwidth,height=5cm]{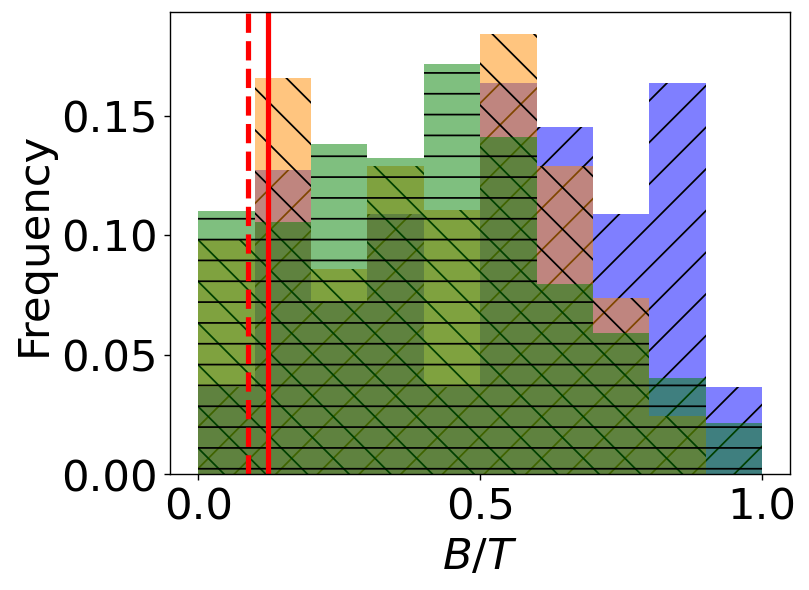}\hfill
    \includegraphics[width=.49\textwidth,height=5.02cm]{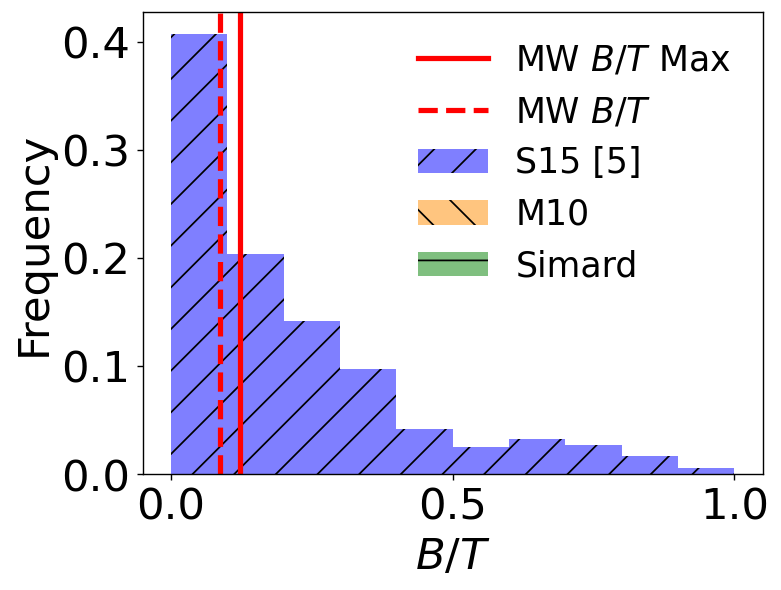}\hfill
    \\[\smallskipamount]
    \caption{Bulge-to-total flux ratio of 55 edge-on (\textit{left}) and 521 non-edge-on (\textit{right}) spiral galaxies from \citetalias{Salo2015} [5]. 163 edge-on galaxies from \citetalias{Mosenkov2010} are also included in the left plot, in addition to $\sim$20,000 edge-on galaxies from \citet{Simard2011}. In each plot, the Milky Way's maximum $B/T$ value from \citetalias{Mosenkov2021} is highlighted, as well as a lower estimate created using the stellar mass ratio of the bulge and bar from \citet{Portail2017}.
    }\label{fig:BT_graphs}
\end{figure*}

\begin{figure*}
    \centering
    \includegraphics[width=0.95\textwidth]{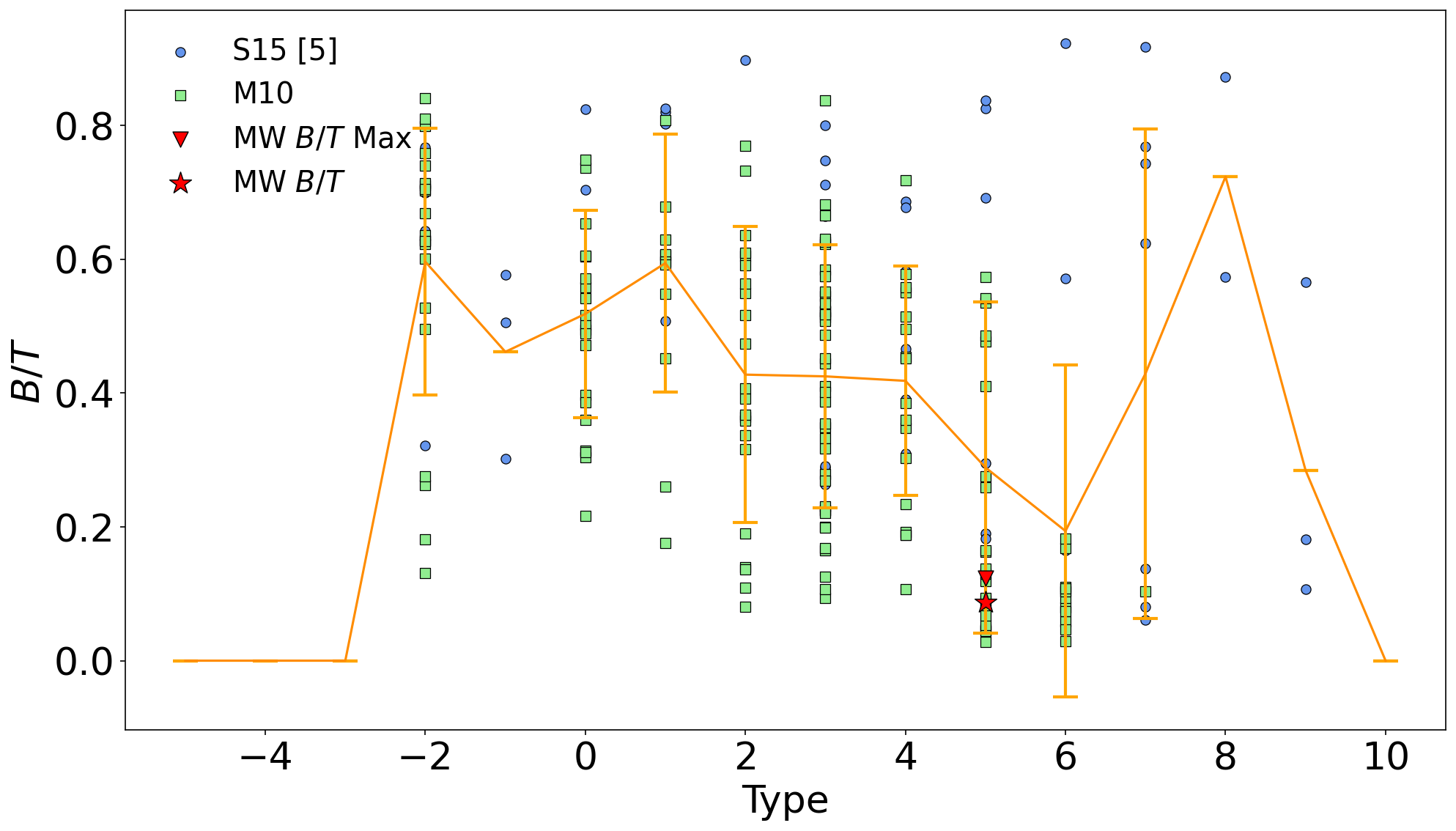}\hfill
    \includegraphics[width=0.95\textwidth]{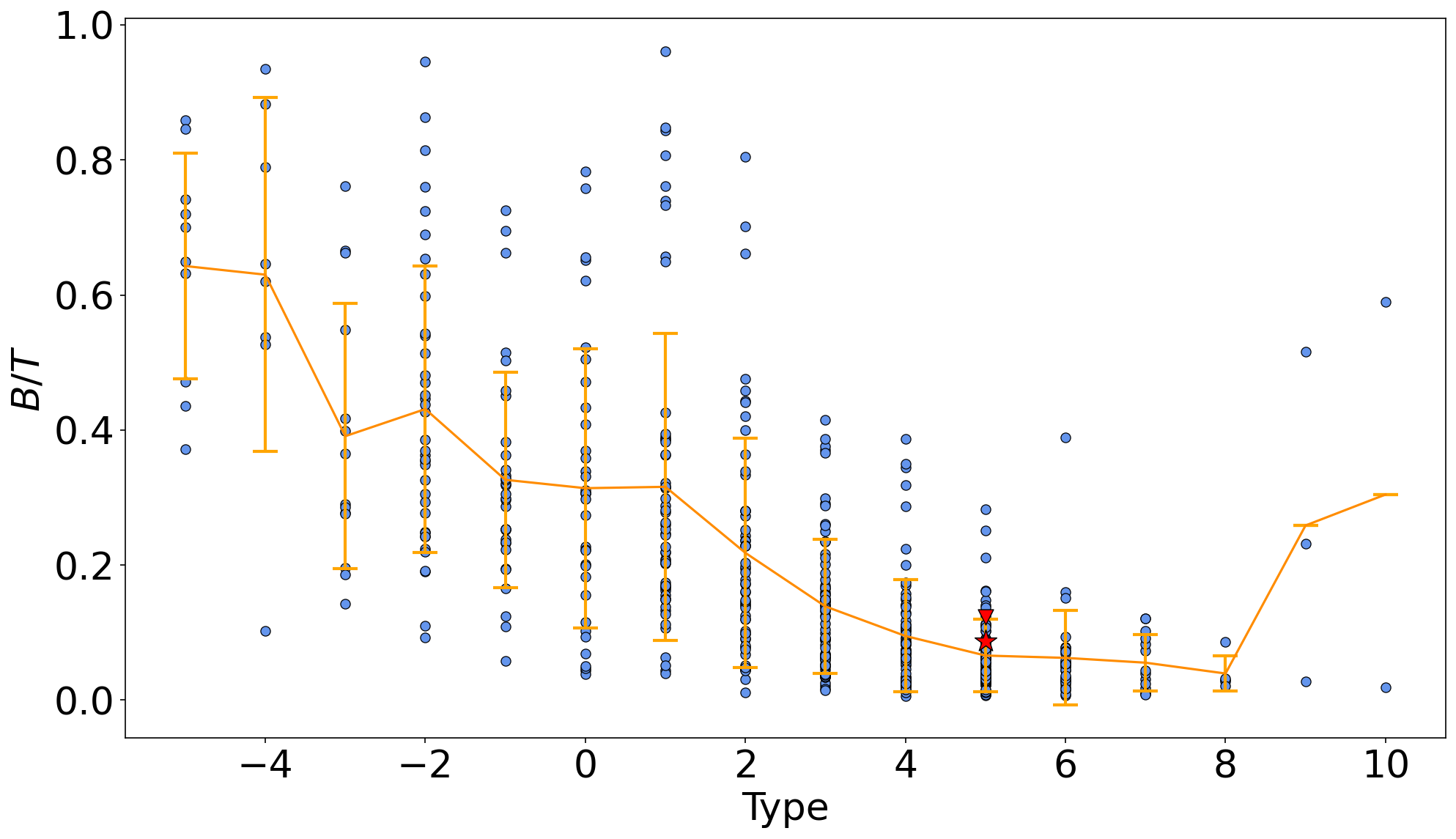}\hfill
    \\[\smallskipamount]
    \caption{Bulge-to-total flux ratio data with the galaxy's numerical Hubble stage from \citet{Sheth2010} (the numerical Hubble stage values are rounded to the nearest integer). In both plots, the line passes through the mean $B/T$ value of each numerical Hubble stage value, with the vertical lines representing one STD above and below the mean. Numerical Hubble stage values with fewer than four data points have no STD calculated. \textit{Top}: 55 edge-on galaxies from \citetalias{Salo2015} [5] and 163 edge-on galaxies from \citetalias{Mosenkov2010}. We utilize a maximum value of 0.12 for the Milky Way's $B/T$ ratio from \citetalias{Mosenkov2021}. We estimate a lower value of 0.09 by using the stellar mass ratio of the bulge and bar from \citet{Portail2017}. We also estimate the Milky Way's numerical Hubble stage as 5, which corresponds to an Sc-type galaxy. Its maximum value data point is highlighted by the red triangle, while the its lower estimate is highlighted by the red star. \textit{Bottom}: 
    521 non-edge-on galaxies from \citetalias{Salo2015} [5] with the same Milky Way data highlighted as in the edge-on galaxy plot.}\label{fig:BT_TT_graphs}
\end{figure*}

Similar to previous sections, the models of galaxies are split between edge-on and non-edge-on galaxies. All of our $B/T$ ratio comparison data for this section comes from \citetalias{Salo2015} and \citetalias{Mosenkov2010}. However, we also compare the $B/T$ ratios of our galaxies with their numerical Hubble stage (Type). For the \citetalias{Salo2015} data, these Type values were gathered from \citet{Sheth2010}, while for the \citetalias{Mosenkov2010} data, they were included in the original paper. Using an Sc classification for the Milky Way \citep{Georgelin1976}, we convert that to a value of 5 for its numerical Type. While Figure~\ref{fig:BT_graphs} compares the Milky Way's $B/T$ value to all the galaxies in our sample, Figure~\ref{fig:BT_TT_graphs} allows us to compare it to other Sc-type galaxies, which is a much better comparison to determine if the Milky Way is abnormal in its $B/T$ value. Nonetheless, we include the comparison to all the galaxies for thoroughness.

We begin our analysis with Figure~\ref{fig:BT_graphs} which displays two histograms of our galaxy data 
(the first histogram shows the edge-on galaxies while the second one shows the non-edge-on galaxies). Unfortunately, the size of the sample of edge-on galaxies is significantly smaller than that of non-edge-on galaxies. This caused the seemingly random pattern for the \citetalias{Salo2015} and \citetalias{Mosenkov2010} galaxies in the left plot. In an attempt to show the true pattern for edge-on galaxies, we include $\sim$20,000 galaxies from \citet{Simard2011}, which we magnitude-restricted to only contain edge-on galaxies of similar absolute magnitude to the Milky Way. From this, we see that edge-on galaxies have seemingly larger bulges than non-edge-on galaxies, with the distribution being much more spread out. While the \citet{Simard2011} galaxies are useful for this visualization, they were obtained through optical observations, so we omit them from the rest of this NIR $B/T$ analysis. Additionally, due to the small size of the sample of edge-on galaxies in the NIR, we will not give much weight to the conclusion in our overall analysis of the Milky Way's bulge.

Looking at the edge-on galaxy histograms in Figure~\ref{fig:BT_graphs} (\textit{left} plot), we see high-frequency bars occurring throughout the $B/T$ ratio for the samples in the NIR, likely due to the presence of bars inflating the $B/T$ value in edge-on galaxies (although we point out that for the Milky Way, the $B/T$ ratio also includes the presence of the bar, hence our effort to correct it) and selection effects. It is clear that the Milky Way's $B/T$ value is far below the average value for these edge-on galaxies. The Q1 cutoff value indicates that the Milky Way's maximum and estimated $B/T$ values are well within the first quarter. Therefore, the Milky Way appears to have an abnormally small $B/T$ value compared to edge-on galaxies in the samples used. If we compare the Milky Way with the sample of $\sim$20,000 edge-on galaxies from \citet{Simard2011}, 
both of the Milky Way's $B/T$ values appear to be somewhat small. However, as previously stated, this comparison is not fair due to the \citet{Simard2011} values being fitted in the optical waveband. Nevertheless, since the $B/T$ ratio is higher in the NIR than in the optical due to the stellar population gradient and differential dust attenuation \citep{2010MNRAS.407..144T,2014A&A...570A...6S,2017A&A...608A..27G,2019MNRAS.483.1862Z,2020AJ....159..195D,2021MNRAS.502.5508P}, we can conclude that the Milky Way's $B/T$ should be quite average among the edge-on galaxies with similar stellar mass.

The histogram for non-edge-on galaxies (\textit{right} plot in Fig.~\ref{fig:BT_graphs}), however, appears very much like how we would expect for late-type galaxies: the frequency rate of galaxies dramatically decreases with increasing $B/T$. The Milky Way's values are well within the range of normality, being between the Q1 cutoff value and the median. 

In Figure~\ref{fig:BT_TT_graphs}, we compare the Milky Way's $B/T$ to morphological type, together with other galaxies in our sample. The Type values given for our dataset originally had various decimal point values, so we rounded these values to the nearest integer for easier visualization. Since our main desire was to view how the $B/T$ values of galaxies with similar Type values compare with one another, this rounding seems acceptable. In each scatter plot, we also represented the mean and standard deviation (STD) values for each grouping. If a grouping had fewer than four galaxies, we disregarded the STD value.

Similar to the histogram in Figure~\ref{fig:BT_graphs} for edge-on galaxies, the \textit{upper} scatter plot in Figure~\ref{fig:BT_TT_graphs} also shows no pattern: the $B/T$ ratio does not correlate with Hubble stage, in contrast to what we see on the \textit{bottom} plot for non-edge-on galaxies. In the bottom plot, the $-$5-Type-value grouping, representing elliptical galaxies, has some of the highest $B/T$ values with the trend then decreasing as the Type values increase, until the $B/T$ values taper off around the 7-8-Type-value groupings\footnote{We note here that despite the negative morphological type for some of the selected galaxies in our samples, their $B/T<1$ values indicate that these galaxies were probably misclassified as elliptical galaxies and should instead class as lenticular galaxies.}.
In the upper plot, the average $B/T$ ratios are typically higher for the same morphological type as compared to the same distribution but for non-edge-on galaxies. The reason for this may be that all morphologies we use in our study are taken from the HyperLeda database \citep{Makarov2014}. Since HyperLeda utilizes optical observations, the morphology of galaxies (especially edge-on galaxies) is heavily affected by dust attenuation, and, thus, may be influenced by the projection effect. Moreover, the bulge contribution, not speaking of the galaxy spiral pattern, is hard to visually determine. At a Type value of 5, both of the Milky Way's $B/T$ values lie within one STD from the mean value for edge-on galaxies, indicating an average $B/T$ value for our Galaxy for its morphological type. Similarly, the Milky Way's maximum $B/T$ value lies just over one STD 
above the mean value for non-edge-on galaxies of the same morphological type, with our lower estimate falling within one STD of the mean. Since our Galaxy's \textit{maximum} $B/T$ value is barely one STD above the mean, we can safely make the assumption that the true value of the Milky Way's $B/T$ ratio is average when compared to non-edge-on galaxies of the same morphology and mass.

\section{Pitch Angle}
\label{sec:pitch_angle}

\begin{figure*}[!t]
    \centering
    \includegraphics[width=0.95\textwidth]{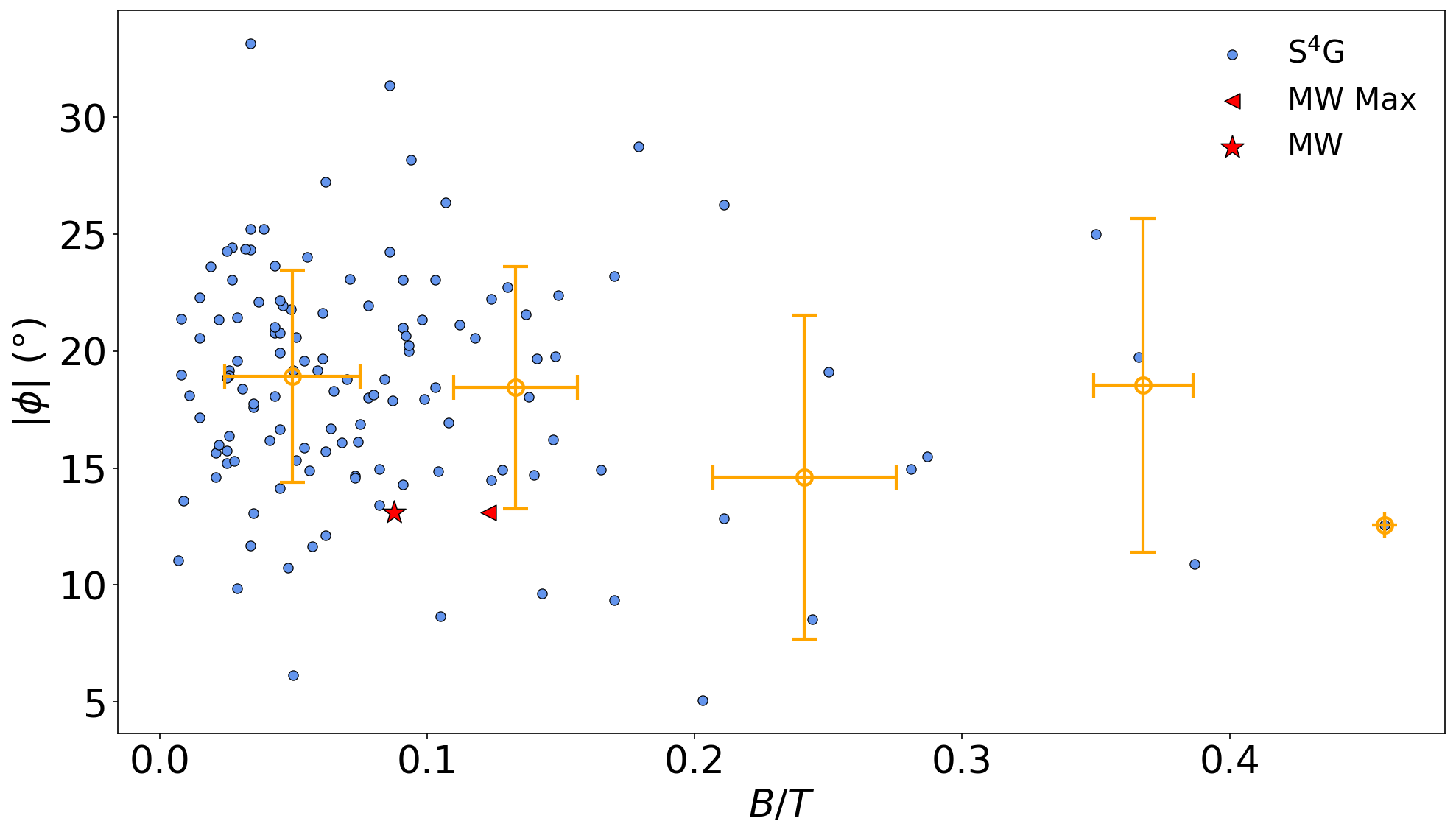}\hfill
    \\[\smallskipamount]
    \caption{127 galaxies with both mean pitch angle data from \citetalias{DiazGarcia2019} and $B/T$ data from \citetalias{Salo2015}. The data was split into five equally-sized bins of $B/T$, whereupon means and STDs of the dataset were calculated. The Milky Way's mean pitch angle value and max $B/T$ value were taken from \citet{Vallee2015} and \citetalias{Mosenkov2021}, respectively, with a lower estimate of the $B/T$ value included (as in Fig.~\ref{fig:BT_TT_graphs}).}\label{fig:pitch_BT}
\end{figure*}

\begin{figure*}[!t]
    \centering
    \includegraphics[width=0.95\textwidth]{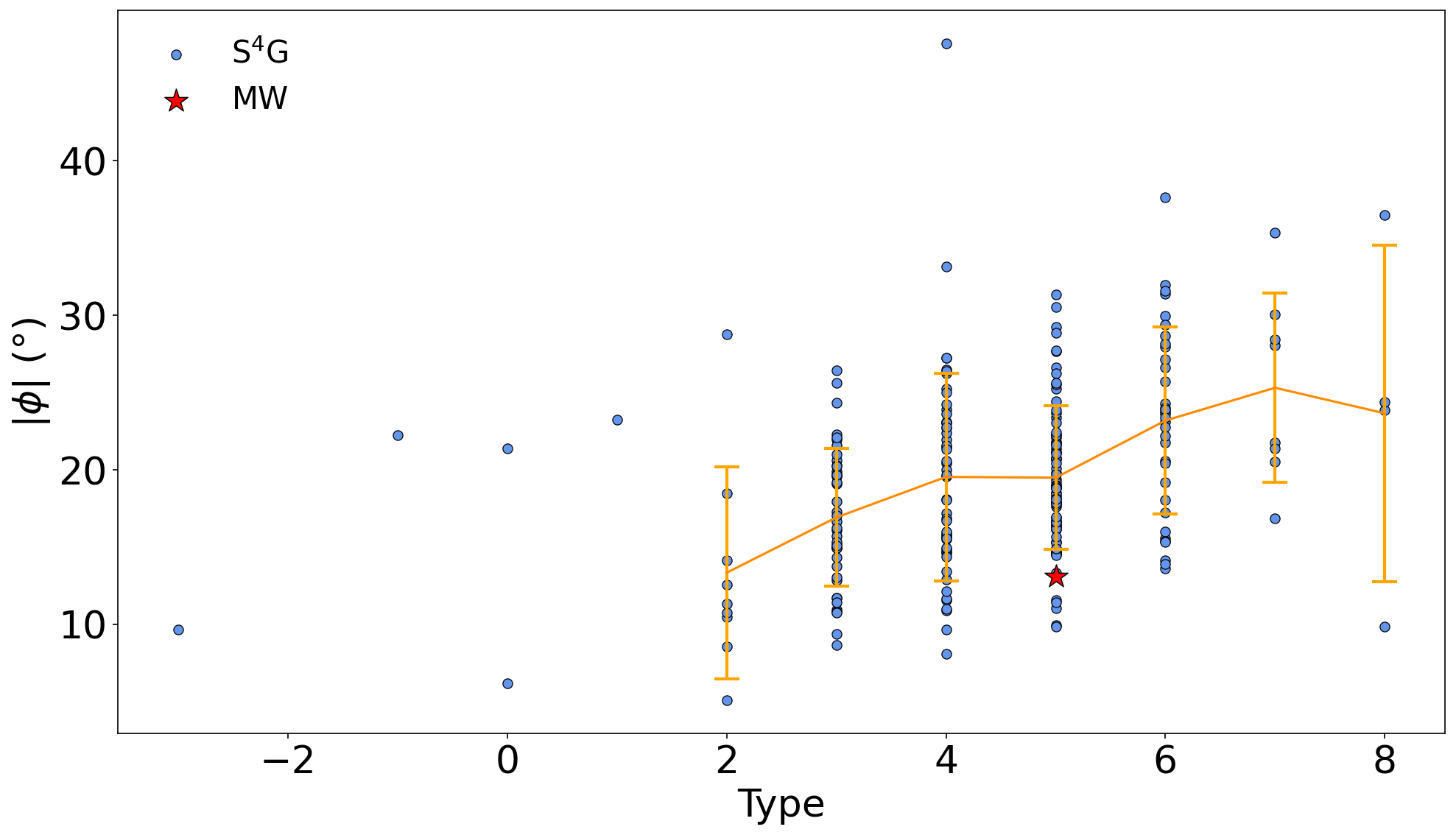}\hfill
    \\[\smallskipamount]
    \caption{Mean pitch angle data from \citetalias{DiazGarcia2019} and rounded numerical Hubble stage data from \citet{Sheth2010}. The line passes through the mean pitch angle value of each numerical Hubble stage value, with one STD being shown. The mean and STD data begins at Type $=2$ due to previous values having too few datapoints. The Milky Way's mean pitch angle value and numerical Hubble stage value were taken from \citet{Vallee2015} and \citet{Georgelin1976}, respectively.}\label{fig:pitch_type}
\end{figure*}

Finally, within this section, we provide a brief analysis of the \textit{mean} pitch angle of the Milky Way compared to galaxies within the S$^4$G survey. For these comparisons, we collected a Milky Way pitch angle from \citet{Vallee2015}, who utilized four different statistical methods to find the mean pitch angle of our Galaxy. Each method produced a result between $12\degree$ and $14\degree$, with the mean of the results being $13.1\degree$. The exact number of the spiral arms in the Milky Way remains uncertain: a two-arm and four-arm spiral pattern has been proposed (see e.g. \citealt{2011ARep...55..108E} and references therein). In a recent study, \citet{2023ApJ...947...54X} conclude that our Galaxy is probably a multiarmed galaxy. Despite this uncertainty on the number of the spiral arms, it is evident that the Milky Way's spiral structure is not flocculent (patchy) because individual arms have been identified, which is an attribute of grand-design and multiarmed galaxies.

For comparison, we utilize the pitch angles for a large sample of multiarmed and grand-design S$^4$G galaxies from \citetalias{DiazGarcia2019}. To aid in this comparison, we used two other parameters: $B/T$ ratios and Type values. For both the Milky Way and external galaxy data, we retrieved these parameter values from the same sources described in Section~\ref{sec:bulge}, although we note that in our $B/T$-pitch-angle comparison, we only use galaxies that have both pitch angle values from \citetalias{DiazGarcia2019} and $B/T$ values from \citetalias{Salo2015}. Additionally, similar to our data in Section~\ref{sec:bulge}, we round the Type value for each galaxy to the nearest integer to better see the trend between earlier- and later-type disk galaxies.

Beginning with Figure~\ref{fig:pitch_BT}, we grouped the data into five equally-sized bins of $B/T$ values before calculating the mean and STD data. It is evident that spiral galaxies are far more likely to have $B/T$ values between 0 and 0.2, as shown by the large number of data points within that range, and so those groupings are more indicative of the actual trend between galactic $B/T$ ratio and pitch angle. Upon grouping the data, we calculated both the mean $B/T$ value and the mean pitch angle for each grouping, along with a standard deviation for each parameter. From these values we see that for galaxies with $B/T$ values between 0.1 and 0.2, the maximum Milky Way value is within the average range of $B/T$ values, though it lies a little more than one STD below the mean pitch angle value. Our lower $B/T$ value estimate falls in the 0.0 to 0.1 $B/T$ category, and is more than one STD above the mean $B/T$ value and more than one STD below the mean pitch angle value for this category. This would imply that the Milky Way's spiral arms are more tightly wound than other galaxies with similar $B/T$ ratios.

Figure~\ref{fig:pitch_type} shows the pitch angle of the S$^4$G galaxies compared to their Type, similar to Figure~\ref{fig:BT_TT_graphs}. Likely due to the same issues with HyperLeda galaxy classification discussed previously (i.e. the galaxies being heavily influenced by dust attenuation because of the observations being performed in the optical waveband), there are a few early-type galaxies in this dataset, though the vast majority are spirals. From this data, we see an interesting trend: as the galaxies progress to later and later Types, the mean pitch angle steadily increases as expected, indicating more loosely wound spiral arms. When we focus on the 5-Type-value grouping, we again see that the Milky Way falls a little more than one STD below the mean, similar to Figure~\ref{fig:pitch_BT}. This would imply that the Milky Way's spiral arms are also more tightly wound than in other galaxies of a similar morphology. To summarize, it appears that when compared to galaxies with similar $B/T$ ratio values and when compared to galaxies with similar morphological types, the Milky Way has more tightly wound spiral arms.

\section{Conclusion}
\label{sec:conclusion}

Studying the structural parameters of our Galaxy is crucial to understanding how the Milky Way fits in with the rest of the Universe. To that end, we have compared various parameters of our Galaxy with other spiral galaxies from a variety of sources. We compared the scale length and scale height of the Milky Way to a large sample of observed disk galaxies using a radiative transfer model in the optical. Also, we performed a similar comparison in the NIR with either a general single-disk or a two-disk (thin+thick disk) model, comparing these models to observed single-disk and two-disk galaxies in a varied sample of edge-on and non-edge-on galaxies. For both optical and NIR comparisons, we've included mass/magnitude versus scale length comparisons to better see the scaling relation of our Galaxy in context. Then, we compared the Milky Way's bulge-to-total luminosity ratio $B/T$ to the entire sample of edge-on and non-edge on galaxies, and also considered the $B/T$ ratio versus Hubble type. Finally, we compared our Galaxy's mean pitch angle to other galaxies with similar $B/T$ values and other galaxies of a similar morphological type. We summarize our findings from these various analyses as follows:

\begin{enumerate}
    \item In the optical, the Milky Way's scale length appears to be below average (falling in the first quarter), while the scale height and flatness ratio for the general stellar disk are very typical of spiral galaxies (falling in the second quarter). 
    \item In the NIR, where the comparison is more straightforward due to a much lower internal extinction, those same structural parameters are all normal for the single-disk models of the Milky Way. Compared to edge-on and non-edge-on galaxies, the Milky Way's values are within either the second or third quarters, depending on the parameter. 
    \item For the two-disk Milky Way models in the NIR, we found that the thick disk has normal scale values when compared to smaller-mass galaxies, though when compared to galaxies of similar mass, both its scale length and scale height are very small, indicating it is thinner and less extended than expected. The thin disk follows a similar, though not quite as extreme, trend: its scale length and scale height are larger than the smaller-mass galaxies, though they still fall short of the Milky Way-mass galaxies. This would imply that while the thick disk of the Milky Way is abnormally small for a galaxy of its mass, the thin disk, while still small, is more how one would expect. 
    \item As for the thick-to-thin disk mass ratio, we considered low (0.43) and high (1.34) estimates, which fall in the first quarter and in the third quarter for the reference sample, respectively. Since we do not know which of these estimates is accurate and only suggest that the true value lies between the two, we cannot state with precision where the true Milky Way disk mass ratio lies. The fact that the low mass ratio estimate falls almost directly on the trendline of Figure~\ref{fig:Mass_scatter_graphs} would seem to imply that, despite the evidence from the linear relationship in Figure~\ref{fig:Mass_ratio_graphs}, a logarithmic relationship shows that the Milky Way's disk mass ratio is larger than normal.
    \item When compared with the full dataset of edge-on galaxies in the NIR, the Milky Way's bulge-to-total flux ratio value falls easily within the first quarter. However, the small size of our sample could have influenced this, and therefore this result should be considered with caution. Compared to non-edge-on galaxies, the Milky Way's bulge-to-total flux ratio value is completely normal, falling comfortably within the second quarter. When only compared to other galaxies of its Hubble type, we received contradicting results. Among edge-on Sc-type galaxies, the Milky Way is within a STD of the mean value. When viewed against non-edge-on Sc-type galaxies, the Milky Way's maximum $B/T$ value is just over one STD above the mean, with our lower estimated value falling within one STD. Given the fact that our sample of non-edge-on galaxies contains far more data points, we consider that comparison more valid. Thus, by attempting to correct for bar corruption in the $B/T$ ratio, we find that the Milky Way's value is well within the range of normal values.
    \item Finally, we found that the absolute value of the mean pitch angle of our Galaxy is below average when compared to galaxies with similar bulge-to-total flux ratio values as well as when compared to galaxies of similar morphological type.
\end{enumerate}

Combining all of these results, we find that the scale length and scale height of the Milky Way in the optical and NIR single-disk models generally lie within the typical range of an average spiral galaxy in the Local Universe. The disagreement between the isothermal and exponential models of the Milky Way in fitting the thick disk indicates that further research may need to be done to determine the correct model for the Milky Way disk. We found that the Milky Way's thick disk and optical radius are unusually small when compared to galaxies of similar mass. This indicates that mergers probably played a less significant role in the formation of the thick disk than in other similar-mass galaxies. For example, the galaxies in the AURIGA cosmological simulations (\citetalias{Pinna2023}) had, on average, 22\% of the thick disk mass come from accretion, which thickened the disk and caused the larger scale height values seen in Figure~\ref{fig:multi_2disk_graphs}. Since the thick disk in our Galaxy is several times thinner than in the \textit{HER}OES (\citetalias{Mosenkov2018}) and AURIGA galaxies, we can conclude that the Milky Way's thick disk formed \textit{in situ} without a significant influence from satellite mergers. Additionally, the Milky Way's mean pitch angle indicates that the spiral arms of the Milky Way are more tightly wound than in other similar galaxies.

Further study in this area could include comparing other structural parameters of the Milky Way, such as the dust disk mass, to other galaxies. Additionally, as further galaxy data is gathered from databases such as RCSEDv2\footnote{https://rcsed2.voxastro.org} \citep{Chilingarian2012,Chilingarian2017}, larger sample sizes (for example, using the recent EGIPS catalog of edge-on galaxies, \citealt{2022MNRAS.511.3063M}) would help give more accurate comparisons for all of the analyses performed in this paper.

\section*{Acknowledgements}

We acknowledge the usage of the HyperLeda database (http://leda.univ-lyon1.fr).

The Spitzer Survey of Stellar Structure in Galaxies (S4G) is a volume-, magnitude-, and size-limited survey of over 2300 nearby galaxies at 3.6 and 4.5{\textmu}m. This is an extremely deep survey reaching an unprecedented 1$\sigma$ surface brightness limit of 3.6{\textmu}m(AB) = 27 mag arcsec$^{-2}$. This translates to a stellar surface density of $<<$ 1 M$_{\odot}$ pc$^{-2}$ ! S4G can thus probe the stellar structure in galaxies in a regime where the gas dominates the stars (typical HI surface density $\sim$ a few M$_{\odot}$ pc$^{-2}$).

This dataset or service is made available by the Infrared Science Archive (IRSA) at IPAC, which is operated by the California Institute of Technology under contract with the National Aeronautics and Space Administration.

\bibliographystyle{raa}
\bibliography{ms2023-0263}

\end{document}